\documentclass[final,3p,times,twocolumn]{elsarticle}

\usepackage{hyperref}
\usepackage{siunitx}
\usepackage{booktabs}
\usepackage{array}
\usepackage{longtable}
\usepackage{ragged2e}
\usepackage{caption}
\usepackage{siunitx}
\usepackage{booktabs}
\usepackage{tabularx}

\usepackage{physics}
\usepackage{amsmath,amssymb,bm}
\usepackage{mathtools}
\usepackage{tabularx}
\usepackage{tikz}
\usetikzlibrary{arrows.meta,positioning,calc,fit,shapes.misc}
\usepackage{subcaption}

\journal{Nuclear Physics B}

\begin{document}

\begin{frontmatter}

%\title{A High-Fidelity Benchmark for Multivariable Control of Outdoor Microalgae Raceway Reactors}
\title{A Comprehensive Benchmark Platform for Process Control Research of Outdoor Microalgae Raceway Reactors}

\author[1]{Enrique Rodríguez-Miranda}
\author[1]{Pablo Otálora}
\author[1]{José González-Hernández}
\author[1]{José Luis Guzmán*}
\author[1]{Manuel Berenguel}

\address[1]{University of Almería, Department of Informatics, CIESOL, ceiA3, La Cañada de San Urbano s/n, Almería, Spain. \\ *Corresponding author: José Luis Guzmán (joseluis.guzman@ual.es)}

%% Abstract
\begin{abstract}
This paper presents a benchmarking framework to evaluate process control strategies in outdoor microalgae raceway reactors, integrating four key control regulation tasks: pH, dissolved oxygen (DO), culture volume through coordinated harvest-dilution actions, and temperature via a sump-mounted spiral heat exchanger. The benchmark is built upon a high-fidelity, experimentally calibrated dynamic model that captures the strongly coupled thermal, physicochemical, and biological processes governing industrial-scale open raceway ponds. A closed-loop simulation environment is provided, featuring realistic actuator constraints, gas transport delays, stiff integration, and a fully specified scenario based on multi-day outdoor disturbances (irradiance, temperature, wind, and humidity). Four user-replaceable controllers define the manipulation of CO$_2$ injection, air bubbling, harvest/dilution sequencing, and heat-exchanger operation. The platform computes a unified global performance index, in addition to individual metrics for each control problem, combining tracking error, gas/energy usage, and biomass productivity, enabling consistent and quantitative comparison of alternative control strategies. Baseline regulatory architectures (On/Off, PI/PID and Economic Model Predictive Control (EMPC)) are included to illustrate the benchmark’s use for classical and advanced control methods. By providing an openly specified, reproducible, and computationally tractable benchmark with well-defined function interfaces, this work aims to bridge control methodology and outdoor algal bioprocess engineering, and to support the development of  multivariable control strategies for disturbance-rich environmental systems.
\end{abstract}
%The model includes detailed representations of radiative transfer, carbonate equilibrium, gas–liquid mass transfer, day/night-dependent oxygen supersaturation, evaporation and meteorology-driven heat exchange, biomass growth kinetics, and the hydraulic behaviour of the sump and paddlewheel zones.


% \begin{highlights}
% \end{highlights}

%% Keywords

\end{frontmatter}

%% \linenumbers  % Opcional

%%%%%%%%%%%%%%%%%%%%%%%%%%%%%%%%%%%%%%%%%%%%%%%%%%%%
% SECTION 1: INTRODUCTION
%%%%%%%%%%%%%%%%%%%%%%%%%%%%%%%%%%%%%%%%%%%%%%%%%%%%
\section{Introduction}
\label{sec:introduction}

Microalgae are photosynthetic microorganisms that have gained notable attention in recent years due to their capacity to contribute simultaneously to environmental sustainability and food-security challenges \citep{severo2025microalgae}. Their fast proliferation under diverse environmental conditions, together with their effectiveness in removing nutrients and contaminants from wastewater streams, positions them as a strong candidate for the development of sustainable bioprocesses \citep{Scenedesmus_wastewater,produccion}. Moreover, as they grow, microalgae assimilate carbon dioxide (CO$_2$) and convert it into cellular biomass, providing a net sink for greenhouse gases and supporting climate-change mitigation efforts \cite{Rodríguez2024}.

Production of microalgae is generally carried out in photobioreactors, which can be broadly categorized into open and closed configurations. Open systems directly harness solar irradiance and enable cost-effective cultivation at large scales, albeit with persistent challenges related to contamination management and the efficiency of biomass recovery. In contrast, closed systems, such as tubular or column photobioreactors, offer improved control over operating conditions, facilitating enhanced productivity and more consistent biomass quality. Selecting between these reactor typologies is typically guided by the desired production scale and end-use application.

Open raceway ponds (ORPs) remain the most cost-effective platform for large-scale microalgae cultivation in the world, and for this reason the benchmark control problem introduced in this work focuses on them. ORPs operation is highly challenged by strong diurnal forcing, weather-driven disturbances, and multi-physics couplings that affect key physicochemical and biological variables \citep{LokeShow2022Globalmarket, Quiroz2025Technoeconomic}. Among these, pH, dissolved oxygen (DO), and culture temperature play central roles in productivity, operational safety, and process reliability. pH and DO reflect the interplay between photosynthesis, respiration, and carbonate chemistry; while temperature modulates growth rates, gas solubility, and light-biomass coupling. As reviewed in~\citep{Guzman2025Microalgaeproduction, Novoveska2023Overviewandchallenges, Nwoba2022Processcontrol, Pires2025Mathematicalmodeling}, recent advances in ORPs modeling and control increasingly exploit first-principles models, hybrid dynamic models, and predictive or learning-based control formulations.

From a control perspective, outdoor raceways constitute a compact yet demanding benchmark system. Their dynamics involve radiative transfer and light attenuation, temperature-dependent carbonate equilibria, nonlinear gas-liquid mass transfer, evaporation and heat losses, and biomass kinetics exhibiting strong day/night regime switching \citep{Fernandez2016Dynamicmodel,Nordio2024Abaco2}. These interacting mechanisms lead to stiff dynamics, actuator saturation (air, CO$_2$, and hydraulic flows), transport delays for gases injected in the sump, and multi-rate disturbances from meteorological variables. Existing studies have explored pH and DO regulation using On/Off classical control, PI/PID implementation, and various Model Predictive Control (MPC)-based approaches~\citep{Caparroz2024Anovel, Pataro2023Alearningbased, RodriguezMiranda2020Diurnalandnocturnal}. Temperature regulation via heat exchangers has also received attention in industrial raceways due to heat losses, evaporative cooling, and thermal inertia \citep{de2017exploiting,RodriguezMiranda2021Indirectregulation}. In parallel, the management of harvest and dilution flows is essential to maintain culture volume, biomass concentration, and productivity under fluctuating environmental conditions \citep{Malek2016Modelinganddynamic, Otalora2024Modelingcontrol}.

Despite these contributions, no openly available benchmark exists that simultaneously captures: (i) pH and DO regulation through CO$_2$ and air injection, (ii) thermal control through a heat exchanger, (iii) hydraulic management via harvest-dilution actuation, and (iv) realistic multi-day outdoor disturbances. Benchmark systems are essential for reproducible research, standardised controller evaluation, and fair comparison across methodologies, following the role historically demanded by the control community~\citep{davison_benchmark_1990} and aligned with current calls for high-fidelity benchmark applications in control engineering~\citep{maestre2025control}.

%\subsection*{Objectives and Contributions}

This work introduces a new benchmark platform for ORPs control that addresses these gaps by integrating an experimentally calibrated, first-principles dynamic model with four fully independent but coupled control tasks:

\begin{itemize}
    \item pH regulation via CO$_2$ injection, driven by carbonate chemistry and photosynthesis-induced alkalinity shifts.
    \item DO regulation through air injection in the reactor sump, compensating day-time oxygen supersaturation and night-time respiration.
    \item Harvest-dilution scheduling to maintain culture depth and biomass concentration, including actuator limits.
    \item Temperature control using a sump-mounted spiral heat exchanger with realistic heat-transfer dynamics and water-side flow actuation.
\end{itemize}

The benchmark incorporates: (i) a thermally coupled raceway model including shortwave and longwave radiative exchange, evaporative cooling, ground conduction, and meteorology; (ii) a macroscopic physicochemical model with gas-liquid mass transfer, day/night-dependent parameters, and total inorganic carbon speciation; (iii) a biological model for irradiance-limited photosynthesis, temperature/pH/DO limitation indices, growth, and respiration; (iv) hydraulic balances with evaporation, inflow, outflow, and volume-dependent dilution; and (v) realistic actuator constraints and gas transport delays. Different performance indexes are defined for each of the independent control problems, that penalize tracking error and resource use (gas and thermal energy), along with several Key Performance Indicators (KPIs) related to the productivity of the system and the sustainability of the operation.

Baseline implementations of On/Off, PI/PID, and MPC controllers are provided to demonstrate the benchmark’s use across classical and advanced architectures. All components feature clear interfaces for user-defined controllers, enabling drop-in comparison and reproducible evaluation under fixed multi-day meteorological scenarios.

Overall, this benchmark offers a transparent and physically grounded platform for process control research in environmental and bioprocess systems, aiming to facilitate methodological progress and bridge the gap between simulation and real ORPs operation.

%%%%%%%%%%%%%%%%%%%%%%%%%%%%%%%%%%%%%%%%%%%%%%%%%%%%
% SECTION 2: MICROALGAE RACEWAY REACTOR MODEL
%%%%%%%%%%%%%%%%%%%%%%%%%%%%%%%%%%%%%%%%%%%%%%%%%%%%
\section{Dynamic model of the microalgae raceway benchmark}
\label{sec:model}
\raggedbottom

\subsection{Physical raceway reactor}
\label{subsec:reactor}

\begin{figure}[!t]
  \centering
  \includegraphics[width=\linewidth]{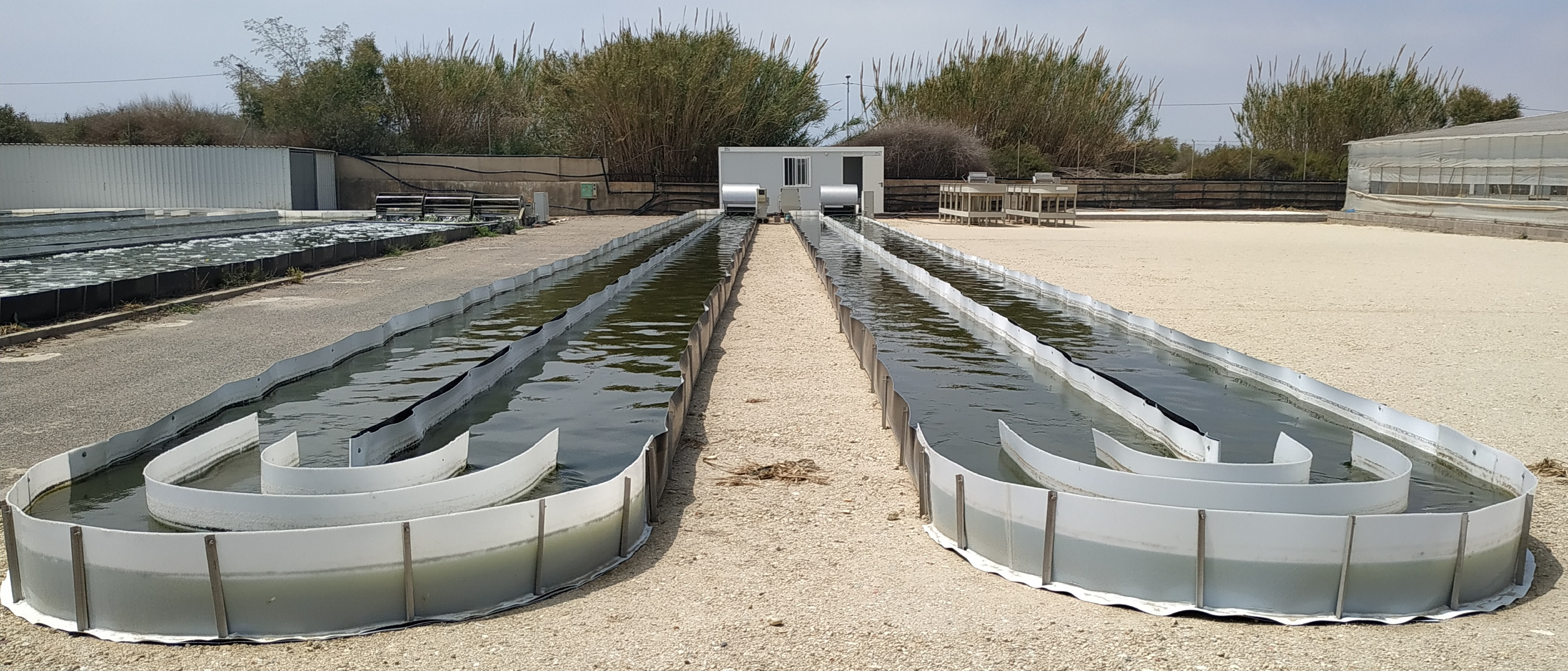}
  \caption{Microalgae raceway reactor at the IFAPA facilities (Almería, Spain).}
  \label{fig:raceway_photo}
\end{figure}

The benchmark is based on a real microalgae raceway reactor belonging to the CIESOL research center from the University of Almer\'ia (UAL), and located at the IFAPA (Instituto de Formación Agraria y Pesquera de Andalucía) research facilities (Figure \ref{fig:raceway_photo}), which are near the UAL campus (Almería, Spain). The reactor consists of two straight channels of length $L=40$~m each and width $W=1$~m, connected by a U-shaped bend, for a total illuminated area of $A = W L = 80$~m$^2$. Mixing and circulation are provided by an aluminum paddlewheel of diameter 1.5~m, driven by an electric motor and operated at a constant superficial liquid velocity of approximately 0.2~m\,s$^{-1}$ via a frequency inverter.

Carbonation takes place in a cylindrical sump of radius $r_{\mathrm{sump}} = 0.4$~m and depth $h_{\mathrm{sump}} = 1.12$~m, located downstream of the paddlewheel. CO$_2$ and air are injected into the sump through plate membrane diffusers fed by two independent gas lines. The raceway walls are built with low-density polyethylene of thickness 3~mm.

The model parameters were calibrated for the strain \textit{Scenedesmus almeriensis}, which is the species typically cultivated in these reactors. The most relevant parameters include a maximum specific growth rate of 0.8 day$^{-1}$, optimal pH and temperature values for growth of 8 and 30$^o$C, respectively, and an upper DO limit of 383.21\% above which growth is inhibited. This values will be tighlty related to the chosen setpoints during the simulation.

\subsection{Model architecture, states and signals}
\label{subsec:states}

The raceway is modeled as a well-mixed tank with explicit representation of sump gas transfer and bulk temperature dynamics. The benchmark implements a closed-loop simulation with a discretization step $T_m = 60$~s and four external controllers (pH, DO, harvest--dilution, and temperature).

The dynamic state vector is
\begin{equation}
  \bm{x}
  =
  \bigl[
    X_{alg},\,
    X_{O_2},\,
    DIC,\,
    Cat,\,
    H,\,
    T,\,
    V
  \bigr]
  \label{eq:state-vector}
\end{equation}
where:
\begin{itemize}
  \item $X_{alg}$ [g\,m$^{-3}$]: biomass concentration.
  \item $X_{O_2}$ [mol\,m$^{-3}$]: dissolved O$_2$.
  \item $DIC$ [mol\,m$^{-3}$]: total inorganic carbon.
  \item $Cat$ [mol\,m$^{-3}$]: strong cation concentration (electroneutrality balance).
  \item $H$ [mol\,m$^{-3}$]: proton concentration.
  \item $T$ [\si{\celsius}]: bulk raceway temperature.
  \item $V$ [m$^3$]: total culture volume.
\end{itemize}

The carbonate species CO$_2$, HCO$_3^-$ and CO$_3^{2-}$ are not states, but are reconstructed algebraically from the carbonate system at each integration step such as described in Section \ref{subsec:engineering}.

The exogenous inputs (disturbances) are global irradiance $\mathrm{RadG}$ [W\,m$^{-2}$], relative humidity $RH$ [\%], ambient temperature $T_{\mathrm{ext}}$ [\si{\celsius}], and wind speed $U_{\mathrm{wind}}$ [m\,s$^{-1}$]. Photosynthetically active radiation (PAR) is computed inside the simulator as
\begin{equation}
  \mathrm{PAR}~[\mu\mathrm{mol\,m^{-2}\,s^{-1}}]
  =
  0.46 \times 4.56 \times \mathrm{RadG}
  \label{eq:PAR-from-RadG}
\end{equation}

The manipulated inputs generated by the controllers are:
\begin{itemize}
  \item $Q_{\mathrm{CO_2}}$ [m$^3$\,s$^{-1}$]: CO$_2$ gas flow rate injected in the sump.
  \item $Q_{\mathrm{air}}$ [m$^3$\,s$^{-1}$]: air flow rate injected in the sump.
  \item $Q_{d_{cmd}}$ [-]: dilution command, that is translated to flow rate internally (fresh medium inlet).
  \item $Q_{h_{cmd}}$ [-]: harvest command, that is translated to flow rate internally (culture outlet).
  \item $Q_w$ [m$^3$\,s$^{-1}$]: water flow through the sump heat exchanger.
  \item $T_{\mathrm{in}}^{\mathrm{HX}}$ [\si{\celsius}]: inlet water temperature in the heat exchanger.
\end{itemize}

The benchmark outputs reported to the user are
\begin{align}
  \mathrm{pH}
  &= -\log_{10}\!\left(\frac{H}{1000}\right)
  \label{eq:pH-def}\\[1pt]
  \mathrm{DO}~[\%]
  &= 100\,
     \frac{X_{O_2}}{X_{O_2}^{eq}(T)}
  \label{eq:DO-def}\\[1pt]
  X_{alg}~[\mathrm{g\,L^{-1}}]
  &= \frac{X_{alg}}{1000}
  \label{eq:Xalg-def}\\[1pt]
  \mathrm{Depth}
  &= \frac{V - V_{\mathrm{sump}}}{W L}
  \label{eq:Depth-def}
\end{align}
where $X_{O_2}^{eq}$ is the saturation concentration given by Henry’s law (Section~\ref{subsec:eq-vanthoff}), and $Depth$ is the culture depth obtained from the volume and geometry, with $V_{\mathrm{sump}}$ corresponding to sump volume.

\begin{figure}[!h]
  \centering
  \includegraphics[width=\linewidth]{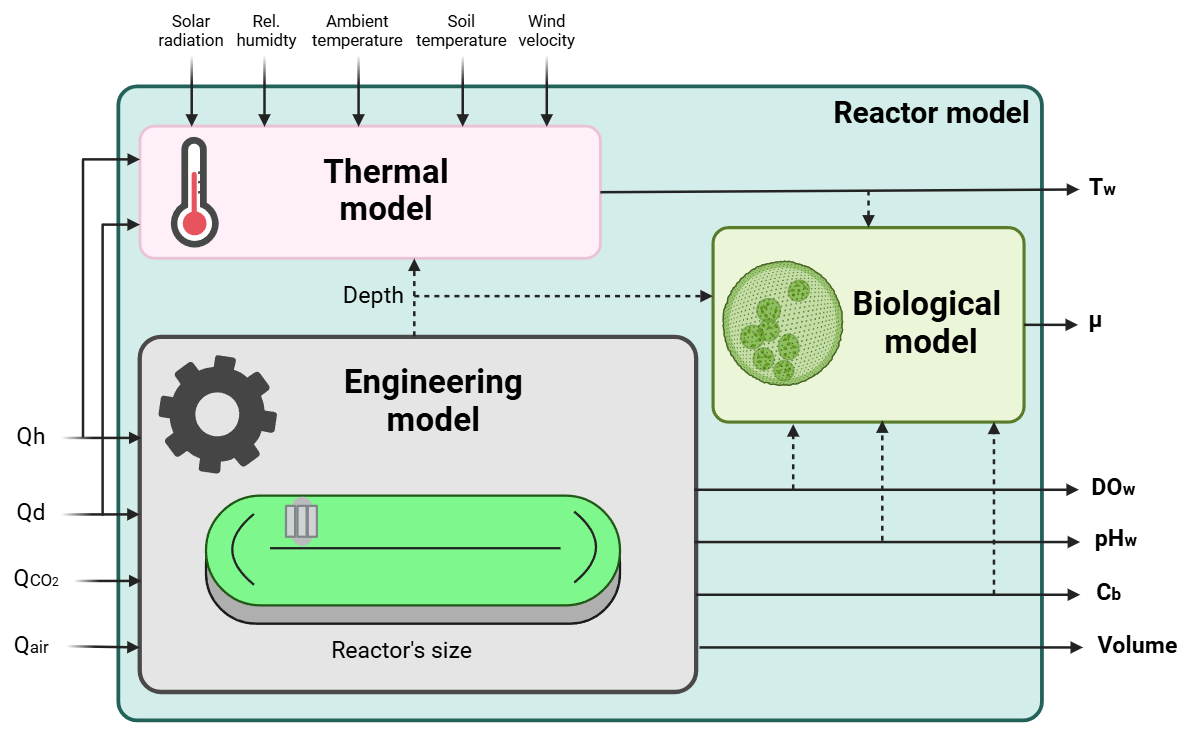}
  \caption{Architecture of the raceway benchmark model. The model is decomposed into three interacting submodels: (i) a \emph{thermal} submodel (energy balance), (ii) a \emph{biological} submodel (microalgae growth) and (iii) an \emph{engineering} submodel (mass transfer, gas transfer, and carbonate system). The three submodels share a common state vector and are driven by meteorological disturbances and actuator signals.}
  \label{fig:model_scheme}
\end{figure}

Figure~\ref{fig:model_scheme} summarizes the three coupled submodels:
\begin{enumerate}
  \item \emph{Thermal submodel}: describes the bulk temperature dynamics including meteorology, evaporation, conduction, convection and the spiral heat exchanger.
  \item \emph{Biological submodel}: computes light, temperature, pH and DO limitation indexes, gross photosynthesis $P$, growth rate $\mu_g$ and maintenance $m$.
  \item \emph{Engineering (macroscopic) submodel}: implements the volume, biomass, DIC, cation and O$_2$ balances, together with gas transfer and carbonate speciation.
\end{enumerate}

All the model parameters and symbols with their corresponding units are summarized in Table \ref{tab:params}.

\subsection{Thermal submodel: energy balance}
\label{subsec:thermal}

The thermal submodel computes the bulk culture temperature $T$ by solving an overall energy balance for the raceway water volume. The formulation follows a previously calibrated temperature model for open raceway reactors \citep{rodriguez-miranda_new_2021}, which was originally developed without active temperature control. In the present benchmark, that model is extended with an additional term describing the heat exchange with a spiral heat exchanger installed in the sump.

The energy balance is written as
\begin{equation}
  \dot{T}
  =
  \frac{Q_{\Sigma}}{\rho_w C_{p,w}\,V}
  -
  \frac{T}{V}\,\dot{V}
  \label{eq:Tdot-thermal}
\end{equation}
where $\rho_w$ and $C_{p,w}$ are the density and specific heat of water, $V$ is the liquid volume, and $\dot{V}$ is given by the hydraulic balance~\eqref{eq:Vdot}. The total heat flow $Q_{\Sigma}$ collects all contributions acting on the well-mixed liquid:
\begin{equation}
\begin{split}
  Q_{\Sigma} =\;
    Q_{\mathrm{irrad}}
  + Q_{\mathrm{rad}}
  + Q_{\mathrm{cond}}
  + Q_{\mathrm{evap}} \\
  + Q_{\mathrm{conv}}
  + Q_{\mathrm{dil}}
  + Q_{\mathrm{harv}}
  + Q_{\mathrm{mix}}
  + Q_{\mathrm{HX}}
\end{split}
\label{eq:Qsum-thermal}
\end{equation}

\paragraph*{Solar irradiance gains ($Q_{\mathrm{irrad}}$)}

The solar heat gain accounts for the fraction of incident global irradiance that is absorbed by the water column:
\begin{equation}
  Q_{\mathrm{irrad}}
  =
  \alpha_{\mathrm{rad}}\,A\,\mathrm{RadG}
  \label{eq:Qirrad-thermal}
\end{equation}
where $\alpha_{\mathrm{rad}}$ is an effective shortwave absorptance of the raceway surface, $A = W L$ is the free-surface area and $\mathrm{RadG}$ is the global solar radiation on the horizontal plane. This term is positive when radiation enters the reactor.

\paragraph*{Longwave radiative exchange ($Q_{\mathrm{rad}}$)}

Longwave exchange represents the net thermal radiation between the raceway surface and the sky:
\begin{equation}
  Q_{\mathrm{rad}}
  =
  \sigma\,\varepsilon_w\,A
  \left(
    T_{\mathrm{sky}}^4 - (T+273.15)^4
  \right)
  \label{eq:Qlw-thermal}
\end{equation}
where $\sigma$ is the Stefan--Boltzmann constant, $\varepsilon_w$ is the longwave emissivity of water and $T_{\mathrm{sky}}$ is an effective sky temperature. The latter is obtained from ambient temperature and humidity using standard clear-sky emissivity relations; thus $Q_{\mathrm{Rad}}$ can be either a net gain or loss depending on the temperature difference.

\paragraph*{Conduction to the ground ($Q_{\mathrm{cond}}$)}

Heat exchange with the ground is modeled as a 1D conduction through the raceway liner and an underlying soil layer:
\begin{align}
  R''
  &=
  \frac{x_{\mathrm{liner}}}{K_{\mathrm{liner}}}
  +
  \frac{x_{ug}}{K_{ug}}
  \label{eq:Rpp-thermal}\\[2pt]
  Q_{\mathrm{cond}}
  &=
  \frac{A}{R''}\,\bigl(T_g - T\bigr)
  \label{eq:Qcond-thermal}
\end{align}
where $x_{\mathrm{liner}}$ and $K_{\mathrm{liner}}$ are the thickness and thermal conductivity of the liner material, $x_{ug}$ and $K_{ug}$ are the thickness and effective thermal conductivity of the underlying ground, and $T_g$ is a ground reference temperature. A positive $Q_{\mathrm{cond}}$ implies heat entering the culture from the ground.

\paragraph*{Evaporation and latent heat ($Q_{\mathrm{evap}}$)}

Evaporation removes mass and latent heat from the raceway surface. The mass evaporation rate is expressed with a linear driving-force formulation:
\begin{equation}
  \dot{m}_e
  =
  k_m\,A\,(C_s - C_a)
  \label{eq:mevap-thermal}
\end{equation}
where $k_m$ is an effective mass-transfer coefficient, and $C_s$ and $C_a$ are the water vapour concentrations at the saturated interface (at temperature $T$) and in the ambient air, respectively. The corresponding latent heat term is
\begin{equation}
  Q_{\mathrm{evap}}
  =
  -\,\ell_v(T)\,\dot{m}_e
  \label{eq:Qevap-thermal}
\end{equation}
where $\ell_v(T)$ is the latent heat of vaporization of water. Since $\dot{m}_e \ge 0$, $Q_{\mathrm{evap}}$ is always a heat loss.

\paragraph*{Convective exchange with air ($Q_{\mathrm{conv}}$)}

Sensible heat exchange with the overlying air is described by a standard convection term:
\begin{equation}
  Q_{\mathrm{conv}}
  =
  h_c\,A\,(T_{\mathrm{ext}} - T)
  \label{eq:Qconv-thermal}
\end{equation}
where $h_c$ is an effective convective heat-transfer coefficient and $T_{\mathrm{ext}}$ is the ambient air temperature. The sign of $Q_{\mathrm{conv}}$ depends on whether the air is warmer or colder than the culture.

\paragraph*{Hydraulic enthalpy flows ($Q_{\mathrm{dil}}$ and $Q_{\mathrm{harv}}$)}

The dilution and harvest flows carry thermal energy in and out of the reactor. Assuming a well-mixed tank, the associated enthalpy terms are
\begin{align}
  Q_{\mathrm{dil}}
  &=
  \rho_w C_{p,w}\,Q_d\,T_{\mathrm{in}}
  \label{eq:Qdil-thermal}\\[2pt]
  Q_{\mathrm{harv}}
  &=
  -\,\rho_w C_{p,w}\,Q_h\,T
  \label{eq:Qharv-thermal}
\end{align}
where $Q_d$ and $Q_h$ are the volumetric flow rates of dilution and harvest, and $T_{\mathrm{in}}$ is the temperature of the incoming medium (here taken equal to ambient temperature). The first term adds enthalpy at $T_{\mathrm{in}}$, while the second removes enthalpy at the current culture temperature.

\paragraph*{Mechanical mixing power ($Q_{\mathrm{mix}}$)}

The paddlewheel introduces mechanical power that is ultimately dissipated as heat. This contribution is modeled as a constant input:
\begin{equation}
  Q_{\mathrm{mix}}
  =
  P_{\mathrm{mix}}
  \label{eq:Qmix-thermal}
\end{equation}
where $P_{\mathrm{mix}}$ is an effective mixing power, calibrated together with the other thermal parameters.

\paragraph*{Spiral heat exchanger in the sump ($Q_{\mathrm{HX}}$)}

The sump contains a spiral tube heat exchanger through which a secondary water stream of flow rate $Q_w$ and inlet temperature $T_{\mathrm{in}}^{\mathrm{HX}}$ circulates. The global heat-transfer coefficient $U$ is defined from the overall thermal resistance between the fluid inside the coil and the raceway water:
\begin{equation}
  \frac{1}{U}
  =
  R_i + R_w + R_o
  \label{eq:U-hx}
\end{equation}
where $R_i$ and $R_o$ are the inner and outer convective resistances, and $R_w$ is the wall conduction resistance. The product $UA$ is obtained as
\begin{equation}
  UA
  =
  U A_o
  \label{eq:UA-hx}
\end{equation}
with $A_o$ the outer surface area of the coil.

Treating the raceway as a large thermal mass and the coil stream as the smaller capacity, the instantaneous heat flow is described by a single-stream NTU formulation:
\begin{align}
  C_w
  &=
  \rho_w C_{p,w}\,Q_w
  \label{eq:Cw-hx}\\[2pt]
  \varepsilon_{\mathrm{HX}}
  &=
  1 - \exp\!\left(-\frac{UA}{\max(C_w,\epsilon)}\right)
  \label{eq:eps-hx}\\[2pt]
  Q_{\mathrm{HX}}
  &=
  C_w\,
  \bigl(T_{\mathrm{in}}^{\mathrm{HX}} - T\bigr)\,
  \varepsilon_{\mathrm{HX}}
  \label{eq:Qhx-thermal}
\end{align}
where $C_w$ is the heat capacity flow rate of the coil water, $\varepsilon_{\mathrm{HX}}$ is the effectiveness of the heat exchanger, and $\epsilon$ is a small regularization constant. A negative $Q_{\mathrm{HX}}$ corresponds to heat removed from the raceway (cooling), whereas a positive value implies heating.

\subsection{Biological submodel: microalgae growth rate}
\label{subsec:bio}

The biological submodel transforms the environmental drivers, photosynthetically active radiation (PAR), temperature, pH and DO, into three main quantities: the gross photosynthetic rate $P$, the specific growth rate $\mu_g$ and the maintenance/respiration rate $m$. This structure follows classical formulations for microalgal growth under coupled light and environmental limitation \citep{Bechet2013, Bernard2012validation}.

\paragraph*{Depth-averaged irradiance and light limitation}

Light limitation is described in terms of a depth-averaged irradiance that accounts for biomass-dependent light attenuation. The mean irradiance is computed from a Beer--Lambert attenuation law:
\begin{align}
  I_{\mathrm{av}}
  &=
  \frac{\mathrm{PAR}}{K_a\,\mathrm{Depth}\,X_{alg}}
  \left(
    1 - \exp\bigl(-K_a\,\mathrm{Depth}\,X_{alg}\bigr)
  \right)
  \label{eq:Iav}\\[2pt]
  \mu_I
  &=
  \frac{I_{\mathrm{av}}^{\,n}}{I_k^{\,n} + I_{\mathrm{av}}^{\,n}}
  \label{eq:muI}
\end{align}
where $K_a$ is the biomass-specific light extinction coefficient, $I_k$ is the constant of irradiance,  and $n$ is a form parameter. The hyperbolic saturation form in \eqref{eq:muI} follows well-known photosynthesis–irradiance curves widely used in raceway modelling \citep{Bechet2013, Jassby1976}. The factor $\mu_I \in [0,1]$ expresses light limitation, with values approaching unity under saturating irradiance.

\paragraph*{Smooth window functions for temperature and pH}

The influence of temperature and pH is captured using smooth window functions that represent a physiological operating range, bounded by a minimum ($a$) and maximum ($c$) value and peaking at an optimum ($b$). For a generic variable $x$ (temperature or pH), the limitation factor is
{\small
\begin{equation}
  \mu_{x}
  =
  \begin{cases}
    0,
    & x \le a \ \text{or}\ x \ge c,\\[2pt]
    3r^2 - 2r^3,
    & a < x \le b,\\[2pt]
    3r^2 - 2r^3,
    & b < x < c,
  \end{cases}
  \qquad
  x \in \{T,\mathrm{pH}\}
  \label{eq:mu-window}
\end{equation}
}
\noindent a cubic Hermite formulation providing smooth transitions across the viable range \citep{Gatamaneni2018Factorsaffecting}. Using calibrated parameter triplets, this yields $\mu_T$ and $\mu_{\mathrm{pH}}$, both within $[0,1]$, with the corresponding maximum ($T_{max}$, $pH_{max}$), minimum ($T_{min}$, $pH_{min}$), and optimal values ($T_{opt}$, $pH_{opt}$), for temperature and pH, respectively.

\paragraph*{DO inhibition of growth}

Accumulation of dissolved oxygen above the equilibrium saturation level inhibits microalgal growth. The DO saturation ratio is
\begin{align}
  \mathrm{DO}
  &= 100\,\frac{X_{O_2}}{X_{O_2}^{eq}(T)}
  \label{eq:DO-level}
\end{align}
and the corresponding inhibition term is
\begin{align}
  \mu_{DO}
  &=
  \!\left(
    1 - \left(\frac{\mathrm{DO}}{(DO_{\max})}\right)^{m_{DO}}
  \right)
  \label{eq:muDO}
\end{align}
following the empirical evidence that high DO supersaturation reduces net photosynthesis and growth due to enhanced photorespiration \citep{Bechet2013}; and where $DO_{\max}$ and $m_{DO}$  are the DO scale for inhibition and the DO inhibition exponent, respectively.

\paragraph*{Gross photosynthesis, growth and maintenance}

The gross photosynthetic rate $P$ results from the multiplicative interaction of all limitation factors:
\begin{align}
  P
  &= \mu_{\max}\,\mu_I\,\mu_T\,\mu_{\mathrm{pH}}\,\mu_{DO}
  \label{eq:P}\\[2pt]
  \mu_g
  &= \eta_X\,P
  \label{eq:mu-g}
\end{align}
where $\mu_{\max}$ is the maximum specific photosynthetic rate and $\eta_X \in [0,1]$ is a dimensionless allocation factor accounting for the fraction of gross photosynthesis effectively converted into specific growth rate ($\mu_g$). Such multiplicative structures are common in dynamic microalgae models for raceways and closed photobioreactors \citep{Bechet2013, Bernard2012validation}.

Maintenance and respiration follow a temperature-dependent formulation:
\begin{align}
  m
  &= m_{\min}
     \Bigl(1 + k_I^{\mathrm{resp}}(1 - \mu_I)\Bigr)
     Q_{10}^{\frac{T-20}{10}}
  \label{eq:m}
\end{align}
where $m_{\min}$ is the basal respiration at \SI{20}{\celsius}, $k_I^{\mathrm{resp}}$ accounts for the increase of respiration under light limitation, and $Q_{10}$ represents the standard temperature sensitivity of metabolic processes \citep{GeiderQ10,LOPEZSANDOVALQ10}. This respiration term ensures a realistic coupling between light availability, metabolic costs and thermal effects.

Together, $P$, $\mu_g$ and $m$ define the biological source and sink terms incorporated into the mass balances for biomass, oxygen and inorganic carbon in the raceway.

\subsection{Engineering submodel: mass transfer, gas transfer, and carbonate system}
\label{subsec:engineering}

The engineering submodel combines hydraulics, biomass mass balance, gas-liquid transfer for CO$_2$ and O$_2$, the dynamics of dissolved inorganic carbon (DIC) and strong cations, and the carbonate system coupled to electroneutrality.

\paragraph*{Hydraulics and effective dilution}

The reactor volume $V$ evolves under dilution, harvesting and evaporation.
\begin{equation}
  \dot{V}
  = Q_d - Q_h - \dot{V}_e
  \label{eq:Vdot}
\end{equation}
with $Q_d$ the dilution inflow, $Q_h$ the harvest outflow and $\dot{V}_e$ the evaporative loss (coupled to the thermal submodel).

\paragraph*{Microalgae biomass balance.}

The biomass concentration $X_{alg}$ obeys the following equation
\begin{equation}
  \dot{X}_{alg}
  =
  (\mu_g - m)\,X_{alg}
  - \frac{Q_d}{V}\,X_{alg}
  \label{eq:Xalg-dot}
\end{equation}
where $\mu_g$ and $m$ are the specific growth and maintenance rates provided by the biological submodel (Section~\ref{subsec:bio}). The first term in the equation corresponds to net biological growth, whereas the second represents biomass washout due to dilution.

\paragraph*{Gas transfer in the sump and at the free surface}

Gas transfer in the sump is described in terms of superficial gas velocities and empirical power-law correlations for the volumetric transfer coefficients. For CO$_2$ injection,
\begin{align}
  U_{g,CO_2}
  &= \frac{Q_{\mathrm{CO_2}}}{A_{\mathrm{sump}}},
  &
  k_La^{CO_2}
  &= \!\bigl(\alpha_{CO_2}\,U_{g,CO_2}^{\beta_{CO_2}}\bigr)
  \label{eq:UgCO2-klaCO2}
\end{align}
where $Q_{\mathrm{CO_2}}$ is the gas flow rate and $A_{\mathrm{sump}}$ is the sump surface area. An analogous expression is used for aeration:
\begin{align}
  U_{g,O_2}
  &= \frac{Q_{\mathrm{air}}}{A_{\mathrm{sump}}},
  &
  k_La^{O_2}
  &= \!\bigl(\alpha_{O_2}\,U_{g,O_2}^{\beta_{O_2}}\bigr)
  \label{eq:UgO2-klaO2}
\end{align}
with $Q_{\mathrm{air}}$ the air flow rate. The effective volumetric transfer coefficients in the well-mixed reactor volume are then
\begin{align}
  k_La^{CO_2,\mathrm{eff}}
  &= k_La^{CO_2}\,\frac{A_{\mathrm{sump}}}{V}
  \label{eq:klaCO2-eff}\\[2pt]
  k_La^{O_2,\mathrm{eff}}
  &= k_La^{O_2}\,\frac{A_{\mathrm{sump}}}{V}
  \label{eq:klaO2-eff}
\end{align}

\iffalse
In addition to sump exchange, gas transfer at the free surface and under the paddlewheel is represented by effective mass-transfer coefficients modulated by the light-dependent factor $\mu_I$ to distinguish day and night operation:
\begin{align}
  k_{CO_2}^{\mathrm{atm,eff}}
  &= k_{CO_2}^{\mathrm{atm}}
     \bigl(\mu_I + (1-\mu_I)\,f_{CO_2}^{\mathrm{atm,night}}\bigr),
  \label{eq:kCO2-atm-eff}\\[2pt]
  k_{CO_2}^{\mathrm{pw,eff}}
  &= k_{CO_2}^{\mathrm{pw}}
     \bigl(\mu_I + (1-\mu_I)\,f_{CO_2}^{\mathrm{pw,night}}\bigr),
  \label{eq:kCO2-pw-eff}\\[2pt]
  k_{O_2}^{\mathrm{atm,eff}}
  &= k_{O_2}^{\mathrm{atm}}
     \bigl(\mu_I + (1-\mu_I)\,f_{O_2}^{\mathrm{atm,night}}\bigr),
  \label{eq:kO2-atm-eff}\\[2pt]
  k_{O_2}^{\mathrm{pw,eff}}
  &= k_{O_2}^{\mathrm{pw}}
     \bigl(\mu_I + (1-\mu_I)\,f_{O_2}^{\mathrm{pw,night}}\bigr),
  \label{eq:kO2-pw-eff}
\end{align}
where $k_{\bullet}^{\mathrm{atm}}$ and $k_{\bullet}^{\mathrm{pw}}$ are the daytime transfer coefficients at the free surface and paddlewheel region, $f_{\bullet}^{\mathrm{night}}$ are night-time reduction factors, and the paddlewheel acts over an effective area
\begin{equation}
  A_{\mathrm{pw}} = W L_{\mathrm{pw}}\mathrm{Depth}.
\end{equation}
\fi

Cross-stripping terms represent the additional transfer of CO$_2$ driven by O$_2$ bubbling and vice versa:
\begin{align}
  k_{CO_2\leftarrow O_2}^{\mathrm{strip,eff}}
  &= k_{\mathrm{strip,CO_2\leftarrow O_2}}\,
     k_La^{O_2,\mathrm{eff}}
  \label{eq:kstrip-CO2}\\[2pt]
  k_{O_2\leftarrow CO_2}^{\mathrm{strip,eff}}
  &= k_{\mathrm{strip,O_2\leftarrow CO_2}}\,
     k_La^{CO_2,\mathrm{eff}}
  \label{eq:kstrip-O2}
\end{align}

\paragraph*{Carbonate speciation}

For given DIC concentration, proton concentration $H$ and temperature-dependent equilibrium constants $K_1(T)$ and $K_2(T)$, the carbonate species are obtained from the standard three-species equilibrium:
\begin{align}
  \Delta
  &= H^2 + H K_1 + K_1 K_2
  \label{eq:Den-carbonates}\\[2pt]
  CO_2
  &= DIC \,\frac{H^2}{\Delta}
  &\\[2pt]
  HCO_3^-
  &= DIC \,\frac{H K_1}{\Delta}
  &\\[2pt]
  CO_3^{2-}
  &= DIC \,\frac{K_1 K_2}{\Delta}
  \label{eq:CO2-HCO3-CO3}\\[2pt]
  OH^-
  &= \frac{K_W}{H}
  \label{eq:OH-from-H}
\end{align}
where $K_W(T)$ is the water autoprotolysis constant. The temperature dependence of $K_1$, $K_2$ and $K_W$ is described in Section~\ref{subsec:eq-vanthoff}.

\paragraph*{DIC and strong cations}

The DIC balance includes dilution, biological uptake and respiration, sump gas transfer, surface exchange and cross-stripping:
\begin{align}
  \dot{DIC}
  &=
  \frac{Q_d}{V}\,(DIC_{\mathrm{in}} - DIC)
  \notag\\[1pt]
  &\quad
  - P\,X_{alg}\,\frac{Y^{\mathrm{alg}}_{CO_2}}{M_{CO_2}}
  + m\,X_{alg}\,\frac{Y^{\mathrm{alg}}_{CO_2}}{M_{CO_2}}
  \notag\\[1pt]
  &\quad
  + k_La^{CO_2,\mathrm{eff}}\,(\tilde{CO}_2^{\mathrm{iny}} - CO_2)
  \notag\\[1pt]
  &\quad
  + k_{CO_2}^{\mathrm{atm}}(CO_2^{eq} - CO_2)
  \notag\\[1pt]
  &\quad
  + k_{CO_2}^{\mathrm{pw}}\,
    \frac{W L_{\mathrm{pw}}\mathrm{Depth}}{V}
    (CO_2^{eq} - CO_2)
  \notag\\[1pt]
  &\quad
  + k_{CO_2\leftarrow O_2}^{\mathrm{strip,eff}}\,(CO_2^{eq} - CO_2)
  \label{eq:DIC-dot}
\end{align}
where $k_{CO_2}^{\mathrm{atm}}$ and $k_{CO_2}^{\mathrm{pw}}$ are the transfer coefficients at the free surface and paddlewheel region, $CO_2^{eq}$ and $\tilde{CO}_2^{\mathrm{iny}}$ are the saturation concentration given by Henry’s law (Section~\ref{subsec:eq-vanthoff}) for CO$_2$ in the atmosphere and injected, and $Y^{\mathrm{alg}}_{CO_2}$ is the stoichiometric yield relating biomass growth and respiration to CO$_2$ consumption and production.

The first line accounts for dilution (inflow and outflow), the second for biological consumption and release of inorganic carbon, the third for sump gas transfer from CO$_2$ injections, and the remaining terms for atmospheric, paddlewheel and cross-stripping exchange with the equilibrium CO$_2$ concentration $CO_2^{eq}$.

Strong cations are assumed conservative except for dilution:
\begin{equation}
  \dot{Cat}
  =
   \frac{Q_d}{V}\,(Cat_{\mathrm{in}} - Cat)
  \label{eq:Cat-dot}
\end{equation}

\paragraph*{Dissolved oxygen}

The dissolved O$_2$ balance includes dilution, biological production and respiration, sump exchange, surface aeration and cross-stripping:
\begin{align}
  \dot{X}_{O_2}
  &=
   \frac{Q_d}{V}\,(X_{O_2}^{eq} - X_{O_2})
  \notag\\[1pt]
  &\quad
  + P\,X_{alg}\,\frac{Y^{\mathrm{alg}}_{O_2}}{M_{O_2}}
  - m\,X_{alg}\,\frac{Y^{\mathrm{alg}}_{O_2}}{M_{O_2}}
  \notag\\[1pt]
  &\quad
  + k_La^{O_2,\mathrm{eff}}(X_{O_2}^{eq} - X_{O_2})
  \notag\\[1pt]
  &\quad
  + k_{O_2}^{\mathrm{atm}}(X_{O_2}^{eq} - X_{O_2})
  \notag\\[1pt]
  &\quad
  + k_{O_2}^{\mathrm{pw}}\,
    \frac{W L_{\mathrm{pw}}\mathrm{Depth}}{V}
    (X_{O_2}^{eq} - X_{O_2})
  \notag\\[1pt]
  &\quad
  + k_{O_2\leftarrow CO_2}^{\mathrm{strip,eff}}\,(- X_{O_2})
  \label{eq:XO2-dot}
\end{align}
Here $X_{O_2}^{eq}$ is the saturation concentration given by Henry’s law, and $Y^{\mathrm{alg}}_{O_2}$ is the stoichiometric yield relating biomass growth and respiration to O$_2$ production and consumption.

\paragraph*{Proton dynamics from electroneutrality}

Electroneutrality is enforced at all times:
\begin{equation}
  f(H,DIC,Cat)
  =
  Cat + H - HCO_3^- - 2\,CO_3^{2-} - OH^-
  = 0
  \label{eq:electroneutrality}
\end{equation}
Differentiating $f=0$ with respect to time and solving for $\dot{H}$ yields
\begin{equation}
  \dot{H}
  =
  -\,\frac{f_{DIC}\,\dot{DIC} + f_{Cat}\,\dot{Cat}}{f_H}
  \label{eq:H-dot-general}
\end{equation}
with partial derivatives
\begin{align}
  f_{DIC}
  &= -\left(\frac{\partial HCO_3^-}{\partial DIC}
           + 2\frac{\partial CO_3^{2-}}{\partial DIC}\right)
  &
  f_{Cat}
  &= 1
  \label{eq:dfdDIC-df dCat}\\[2pt]
  f_H
  &= 1 - \frac{\partial HCO_3^-}{\partial H}
        - 2\frac{\partial CO_3^{2-}}{\partial H}
        + \frac{K_W}{H^2}
  \label{eq:dfdH}
\end{align}
Using the algebraic expressions~\eqref{eq:CO2-HCO3-CO3}, it is convenient to introduce the ratios
\begin{equation}
  g = \frac{HCO_3^-}{DIC}\qquad
  h = \frac{CO_3^{2-}}{DIC}
\end{equation}
which give
\begin{align}
  \frac{\partial HCO_3^-}{\partial DIC}
  &= g
  &
  \frac{\partial CO_3^{2-}}{\partial DIC}
  &= h
  \label{eq:dHCO3dDIC-dCO3dDIC}\\[2pt]
  \frac{\partial HCO_3^-}{\partial H}
  &= DIC\,\frac{\partial g}{\partial H}
  &
  \frac{\partial CO_3^{2-}}{\partial H}
  &= DIC\,\frac{\partial h}{\partial H}
  \label{eq:dHCO3dH-dCO3dH}
\end{align}
with
\begin{align}
  \frac{\partial \Delta}{\partial H}
  &= 2H + K_1
  \label{eq:dDendH}\\[2pt]
  \frac{\partial g}{\partial H}
  &= \frac{K_1(K_1K_2 - H^2)}{\Delta^2}
  &
  \frac{\partial h}{\partial H}
  &= -\,\frac{K_1 K_2}{\Delta^2}\,\frac{\partial \Delta}{\partial H}
  \label{eq:dgdH-dhdH}
\end{align}
one obtains explicit expressions for $f_{DIC}$ and $f_H$ and therefore for $\dot{H}$ through~\eqref{eq:H-dot-general}.

\subsection{Gas-liquid equilibria and Van’t~Hoff temperature dependence}
\label{subsec:eq-vanthoff}

Gas-liquid equilibrium concentrations for O$_2$ and CO$_2$ and the temperature dependence of the carbonate equilibrium constants are described using Henry’s law and Van’t~Hoff-type relations.

\paragraph*{Henry’s law for O\texorpdfstring{$_2$}{O2} and CO\texorpdfstring{$_2$}{CO2}}

Gas--liquid equilibrium is described explicitly for oxygen in equilibrium with atmospheric air, carbon dioxide in equilibrium with atmospheric air, and carbon dioxide associated with the injection of pure CO$_2$. This formulation allows distinguishing between atmospheric exchange and gas injection mechanisms within the reactor.
\begin{align}
  X_{O_2}^{eq}(T)
  &=
  K_H^{O_2}(T)\,p_{\mathrm{atm}}\,y_{O_2} \\[2pt]
  CO_2^{eq}(T)
  &=
  K_H^{CO_2}(T)\,p_{\mathrm{atm}}\,y_{CO_2} \\[2pt]
  \tilde{CO}_2^{\mathrm{iny}}(T)
  &=
  K_H^{CO_2}(T)\,p_{\mathrm{atm}}\,y_{pure_{CO_2}}
  \label{eq:Henry-gas}
\end{align}
Here, $X_{O_2}^{eq}$ and $CO_2^{eq}$ [mol\,m$^{-3}$] denote the equilibrium dissolved concentrations of oxygen and carbon dioxide in contact with atmospheric air, respectively, while $\tilde{CO}_2^{\mathrm{iny}}$ [mol\,m$^{-3}$] represents the equilibrium dissolved concentration associated with the injection of pure carbon dioxide.  
$K_H^{O_2}(T)$ and $K_H^{CO_2}(T)$ [mol\,m$^{-3}$\,atm$^{-1}$] are the temperature-dependent Henry constants for O$_2$ and CO$_2$, $p_{\mathrm{atm}}$ [atm] is the atmospheric pressure, $y_{O_2}$ and $y_{CO_2}$ are the molar fractions of O$_2$ and CO$_2$ in air, and $y_{\mathrm{CO_2}}^{\mathrm{iny}}=1$ corresponds to pure CO$_2$ injection.

For a generic gas $G\in\{O_2,CO_2\}$ in contact with the atmosphere the temperature dependence of the Henry constant is modeled as
\begin{equation}
  K_H^G(T)
  =
  K_{H,\mathrm{ref}}^G\,
  \exp\!\left(
    C_G\left(\frac{1}{T} - \frac{1}{T_{\mathrm{ref}}}\right)
  \right)
  \label{eq:KH-vanthoff}
\end{equation}
with $K_{H,\mathrm{ref}}^G$ the reference value at $T_{\mathrm{ref}} = 298.15$~K and $C_G$ a fitted constant.

\paragraph*{Carbonate equilibrium constants}

The first and second dissociation constants of carbonic acid and the water autoprotolysis constant are given Van’t~Hoff-type dependencies:
\begin{align}
  K_1(T)
  &=
  K_{1,\mathrm{ref}}\,
  \exp\!\left(
    -\frac{\Delta H_{K_1}}{R}
    \Bigl(\frac{1}{T}-\frac{1}{T_{\mathrm{ref}}}\Bigr)
  \right)\\[2pt]
  K_2(T)
  &=
  K_{2,\mathrm{ref}}\,
  \exp\!\left(
    -\frac{\Delta H_{K_2}}{R}
    \Bigl(\frac{1}{T}-\frac{1}{T_{\mathrm{ref}}}\Bigr)
  \right)\\[2pt]
  K_W(T)
  &=
  K_{W,\mathrm{ref}}\,
  \exp\!\left(
    -\frac{\Delta H_{K_W}}{R}
    \Bigl(\frac{1}{T}-\frac{1}{T_{\mathrm{ref}}}\Bigr)
  \right)
  \label{eq:K1-K2-KW-vanthoff}
\end{align}
where $K_{1,\mathrm{ref}}$, $K_{2,\mathrm{ref}}$ and $K_{W,\mathrm{ref}}$ are reference values at $T_{\mathrm{ref}}$, $\Delta H_{K_1}$, $\Delta H_{K_2}$ and $\Delta H_{K_W}$ are effective enthalpies, and $R$ is the universal gas constant. These temperature-dependent constants enter the carbonate speciation and electroneutrality equations described in Section~\ref{subsec:engineering}.

\subsection{Model validation}
\label{subsec:validation}

The model was calibrated with a dataset of ten consecutive days using real data from the raceway reactor described in Section \ref{subsec:reactor}. Afterwards, a validation stage was conducted using an independent dataset covering six consecutive days of operation. The aim was to assess the capability of the calibrated dynamic model to reproduce the behaviour of the raceway reactor under realistic environmental forcing, aeration inputs, and biological activity. The validate dataset includes measurements of pH, DO, reactor temperature, gas injection flows (CO$_2$ and air), solar radiation, external temperature, relative humidity, and wind speed. These variables provide a sufficiently rich excitation of the system to 
challenge the model in both day-night transitions and rapid transients induced by gas injection pulses.

\begin{figure*}[!t]
  \centering
  \includegraphics[width=\linewidth]{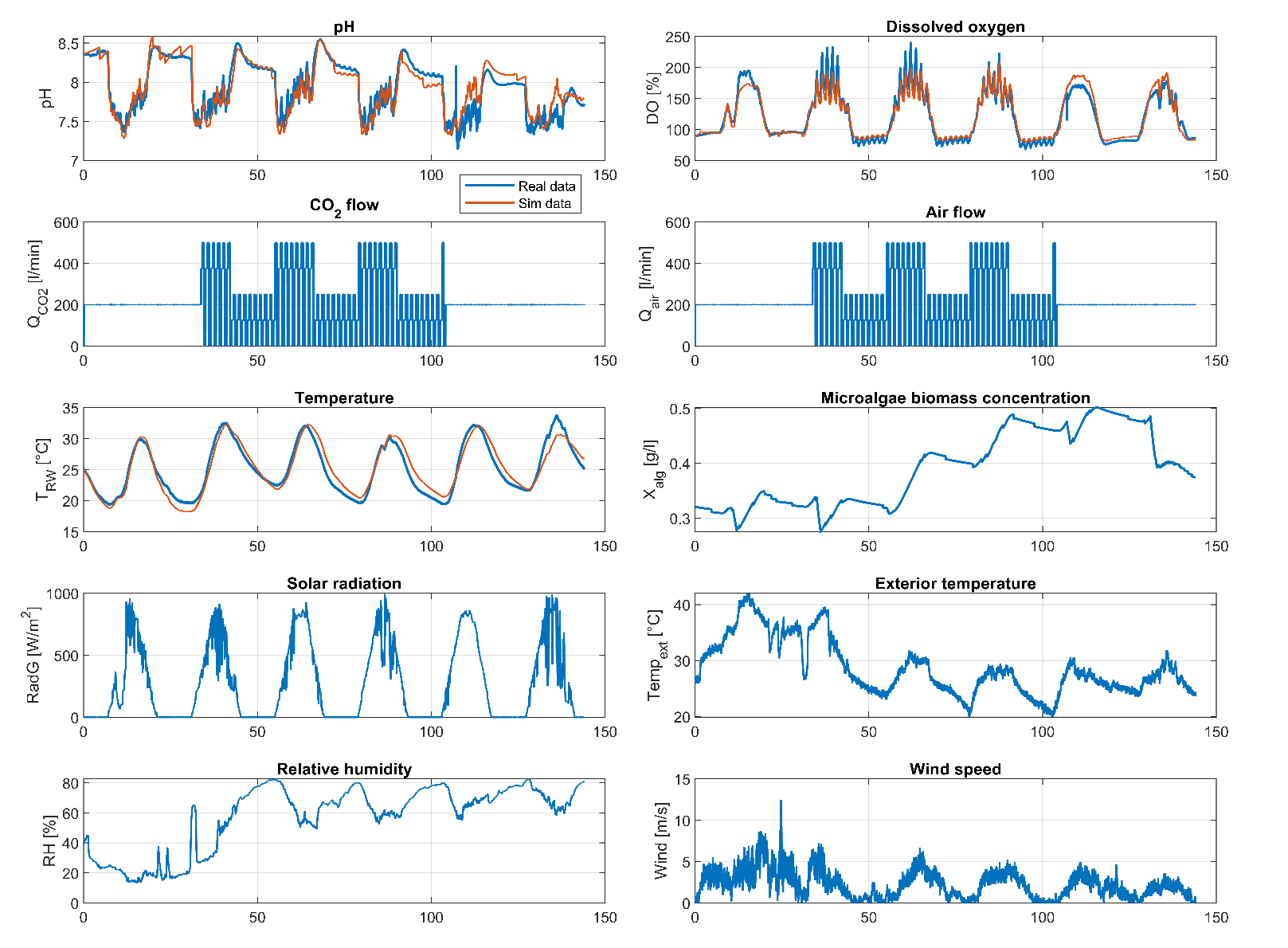}
    \caption{Model validation over a six-day period. Comparison between simulation (solid lines) and 
    measurements (dashed lines) for pH, dissolved oxygen (DO), CO$_2$ injection, air flow, reactor temperature, 
    microalgae biomass concentration, solar radiation, external temperature, relative humidity, and wind speed (REVISAR LEYENDAS).}
    \label{fig:validation-results}
\end{figure*}

Figure~\ref{fig:validation-results} shows the comparison between simulation and measurements for all relevant process variables. The model reproduces accurately the daily thermal cycle, the evolution of pH, and the main 
dynamics of DO, including the characteristic increase during high irradiance periods and the oscillations caused by air injection. The simulated microalgae concentration exhibits the expected biomass growth trend consistent with the six-day horizon in which it can be seen that on the third and fourth day no harvesting and dilution operations are carried out (corresponding to the weekend), causing an increase in concentration. The agreement between model and data across multiple variables demonstrates that the parameters obtained during calibration generalise well beyond the identification window.

Overall, the results indicate that the coupled biological-chemical-thermal model captures the essential dynamics of the open raceway system, including the interaction between CO$_2$ addition, oxygen saturation, 
and pH shifts. Remaining discrepancies are mostly associated with short-timescale sensor noise and transient aeration pulses of high frequency, which are difficult to reproduce exactly in a deterministic 
framework. Nevertheless, the predictive performance is sufficient for control benchmarking and scenario analysis.

% --- Full-width parameter table in single-column mode ---
% \onecolumn
Table~\ref{tab:params} compiles all parameters appearing in the thermal,
biological, and engineering submodels, together with their description and
units, providing a self-contained reference for the benchmark implementation.

% table_parameters_no_values.tex
\begingroup
\sisetup{
  detect-all,
  table-number-alignment = center
}

\begin{table*}[!h]
\centering
\scriptsize
\caption{Model parameters: symbols, descriptions, and units.}
\label{tab:params}
\begin{tabularx}{\textwidth}{@{}lXc lXc@{}}
\toprule
\textbf{Symbol / Name} & \textbf{Description} & \textbf{Unit} &
\textbf{Symbol / Name} & \textbf{Description} & \textbf{Unit} \\
\midrule

% ------------------------------------------------------------------
\multicolumn{6}{@{}l}{\textbf{Main variables}}\\
% Strict order: Cat, CO2, CO3, Depth, DIC, H, HCO3, T, V, X_alg, X_O2

$Cat$ & Strong cation concentration & \si{mol.m^{-3}} &
$HCO_3^-$ & Dissolved bicarbonate concentration & \si{mol.m^{-3}}\\

$CO_2$ & Dissolved CO$_2$ concentration & \si{mol.m^{-3}} &
$T$ & Raceway temperature & \si{\celsius}\\

$CO_3^{2-}$ & Dissolved carbonate concentration & \si{mol.m^{-3}} &
$V$ & Total culture volume & \si{m^3}\\

$Depth$ & Raceway depth & \si{m} &
$X_{alg}$ & Microalgae biomass concentration & \si{g.m^{-3}}\\

$DIC$ & Dissolved inorganic carbon concentration & \si{mol.m^{-3}} &
$X_{O_2}$ & Oxygen concentration & \si{mol.m^{-3}}\\

$H$ & Proton concentration & \si{mol.m^{-3}} & & & \\

% ------------------------------------------------------------------
\addlinespace[0.4em]
\multicolumn{6}{@{}l}{\textbf{Manipulated inputs}}\\
% Strict order: Q_air, Q_CO2, Q_d_cmd, Q_h_cmd, Q_w, T_in^HX

$Q_{\mathrm{air}}$ & Air flow rate injected in the sump & \si{m^{3}.s^{-1}} &
$Q_{h_{cmd}}$ & Harvest command & --\\

$Q_{\mathrm{CO_2}}$ & CO$_2$ gas flow rate injected in the sump & \si{m^{3}.s^{-1}} &
$Q_w$ & Water flow through the sump heat exchanger & \si{m^{3}.s^{-1}}\\

$Q_{d_{cmd}}$ & Dilution command & -- &
$T_{\mathrm{in}}^{\mathrm{HX}}$ & HX inlet water temperature & \si{\celsius}\\

% ------------------------------------------------------------------
\addlinespace[0.4em]
\multicolumn{6}{@{}l}{\textbf{Disturbances}}\\
% Strict order: PAR, RadG, RH, T_ext, U_wind

$PAR$ & Photosynthetically active radiation & \si{\micro\mole.m^{-2}.s^{-1}} &
$T_{ext}$ & Ambient temperature & \si{\celsius}\\

$RadG$ & Global solar irradiance & \si{W.m^{-2}} &
$U_{wind}$ & Wind speed & \si{m.s^{-1}}\\

$RH$ & Relative humidity & \si{\percent} & & & \\

% ------------------------------------------------------------------
\addlinespace[0.4em]
\multicolumn{6}{@{}l}{\textbf{Thermal submodel}}\\
% Latins (strict): A, A_o, C_a, C_{p,w}, C_s, C_w, D_o, h_c, k_m, K_liner, K_ug, L, L_HX, \dot{m}_e, P_mix, Pr_w, Q_d, Q_h
% then: R_i, R_o, R_w, T_g, T_in, T_sky, U, W, x_liner, x_ug
% Greeks after latins (strict): \alpha_rad, \epsilon, \varepsilon_HX, \varepsilon_w, \ell_v(T), \rho_w, \sigma
% Total 35 -> left 18, right 17 (last row right blank)

$A=WL$ & Raceway reactor surface & \si{m^2} &
$R_i$ & HX inner convective resistance & \si{m^2.K.W^{-1}}\\

$A_o=\pi D_o L_{\mathrm{HX}}$ & HX external area (derived) & \si{m^2} &
$R_o$ & HX outer convective resistance & \si{m^2.K.W^{-1}}\\

$C_a$ & Water vapour concentration at air temperature & \si{kg.m^{-3}} &
$R_w$ & HX wall conduction resistance & \si{m^2.K.W^{-1}}\\

$C_{p,w}$ & Water specific heat & \si{J.kg^{-1}.K^{-1}} &
$T_g$ & Ground reference temperature & \si{\celsius}\\

$C_s$ & Vapour concentration at saturated interface & \si{kg.m^{-3}} &
$T_{in}$ & Inlet water temperature & \si{\celsius}\\

$C_w$ & Heat capacity flow rate of the coil & \si{W.K^{-1}} &
$T_{sky}$ & Effective sky temperature & \si{\celsius}\\

$D_o$ & HX outer diameter & \si{m} &
$U$ & Global heat-transfer coefficient (HX) & \si{W.m^{-2}.K^{-1}}\\

$h_c$ & Convective heat-transfer coefficient & \si{W.m^{-2}.K^{-1}} &
$W$ & Raceway width & \si{m}\\

$k_m$ & Mass-transfer coefficient for evaporation & \si{m.s^{-1}} &
$x_{\mathrm{liner}}$ & Liner thickness & \si{m}\\

$K_{\mathrm{liner}}$ & Liner conductivity & \si{W.m^{-1}.K^{-1}} &
$x_{ug}$ & Subgrade thickness (effective) & \si{m}\\

$K_{ug}$ & Subgrade conductivity (effective) & \si{W.m^{-1}.K^{-1}} &
$\alpha_{\mathrm{rad}}$ & Shortwave absorptance of water & --\\

$L$ & Raceway length & \si{m} &
$\epsilon$ & Regularization constant & --\\

$L_{\mathrm{HX}}$ & HX length & \si{m} &
$\varepsilon_{\mathrm{HX}}$ & Heat exchanger effectiveness & --\\

$\dot{m}_e$ & Mass evaporation rate & \si{kg.s^{-1}} &
$\varepsilon_{\mathrm{w}}$ & Water emissivity & --\\

$P_{\mathrm{mix}}$ & Mixing power injected as heat & \si{W} &
$\ell_v(T)$ & Latent heat of vaporization & \si{J.kg^{-1}}\\

$Pr_w$ & Prandtl number (coil water) & -- &
$\rho_w$ & Water density & \si{kg.m^{-3}}\\

$Q_d$ & Dilution flow & \si{m^3.s^{-1}} &
$\sigma$ & Stefan--Boltzmann constant & \si{W.m^{-2}.K^{-4}}\\

$Q_h$ & Harvest flow & \si{m^3.s^{-1}} & & & \\

% ------------------------------------------------------------------
\addlinespace[0.4em]
\multicolumn{6}{@{}l}{\textbf{Biological submodel}}\\
% Latins (strict): DO, DO_max, I_av, I_k, K_a, k_resp^I, m, m_DO, m_min, n, P, pH_max, pH_min, pH_opt, Q10, T_max, T_min, T_opt, X_O2^eq
% Greeks after: \eta_X, \mu_g, \mu_I, \mu_{\max}, \mu_{pH}, \mu_T
% Total 25 -> left 13, right 12 (last row right blank)

$DO$ & Dissolved oxygen & \si{\percent} &
$\mathrm{pH}_{\mathrm{opt}}$ & pH window optimum & --\\

$DO_{\max}$ & DO inhibition scale & \si{\percent} &
$Q_{10}$ & Temperature factor (maintenance) & --\\

$I_{av}$ & Mean irradiance & \si{\micro\mole.m^{-2}.s^{-1}} &
$T_{\max}$ & Temperature window maximum & \si{\celsius}\\

$I_k$ & PAR half-saturation constant & \si{\micro\mole.m^{-2}.s^{-1}} &
$T_{\min}$ & Temperature window minimum & \si{\celsius}\\

$K_a$ & Optical attenuation coefficient & \si{m^2.g^{-1}} &
$T_{\mathrm{opt}}$ & Temperature window optimum & \si{\celsius}\\

$k^{\mathrm{resp}}_I$ & Low-light maintenance gain & -- &
$X_{O_2}^{eq}$ & Oxygen equilibrium concentration & \si{mol.m^{-3}}\\

$m$ & Maintenance and respiration rate & \si{s^{-1}} &
$\eta_X$ & Biomass conversion efficiency & --\\

$m_{DO}$ & DO inhibition exponent & -- &
$\mu_g$ & Specific growth rate & \si{s^{-1}}\\

$m_{\min}$ & Basal maintenance rate & \si{s^{-1}} &
$\mu_I$ & Light limitation factor & --\\

$n$ & Light-response Hill exponent & -- &
$\mu_{\max}$ & Maximum gross photosynthesis rate & \si{s^{-1}}\\

$P$ & Gross photosynthetic rate & \si{s^{-1}} &
$\mu_{pH}$ & pH limitation factor & --\\

$\mathrm{pH}_{\max}$ & pH window maximum & -- &
$\mu_T$ & Temperature limitation factor & --\\

$\mathrm{pH}_{\min}$ & pH window minimum & -- & & & \\

% ------------------------------------------------------------------
\addlinespace[0.4em]
\multicolumn{6}{@{}l}{\textbf{Engineering submodel}}\\
% Latins (strict, \dot{V}_e as V): A_sump, C_G, Cat_in, DIC_in, h_sump,
% k_La^{CO2}, k_La^{O2}, k_atm_CO2, k_atm_O2, k_pw_CO2, k_pw_O2,
% k_strip_CO2<-O2, k_strip_O2<-CO2, K1_ref, K2_ref, K_H^{CO2}, K_H^{O2}, K_Href^G, K_Wref,
% M_CO2, M_O2, p_atm, r_sump, U_g,CO2, U_g,O2, V_sump, \dot{V}_e,
% Y_alg_CO2, Y_alg_O2, y_CO2, y_G, y_O2, y_pureCO2
% Greeks after: \alpha_{CO2}, \alpha_{O2}, \beta_{CO2}, \beta_{O2}, \Delta H_{K_1}, \Delta H_{K_2}, \Delta H_{K_W}
% Total 40 -> left 20, right 20 (no blanks)

$A_{\mathrm{sump}}=\pi r_{\mathrm{sump}}^2$ & Sump area (derived) & \si{m^2} &
$M_{O_2}$ & O$_2$ molar mass & \si{g.mol^{-1}}\\

$C_G$ & Temperature factor for Henry's law & -- &
$p_{\mathrm{atm}}$ & Atmospheric pressure & \si{atm}\\

$Cat^+_{\mathrm{in}}$ & Inlet strong cations & \si{mol.L^{-1}} &
$r_{\mathrm{sump}}$ & Sump radius & \si{m}\\

$DIC_{\mathrm{in}}$ & Inlet DIC & \si{mol.L^{-1}} &
$U_{g,CO_2}$ & Superficial gas velocity (CO$_2$) & \si{m.s^{-1}}\\

$h_{\mathrm{sump}}$ & Sump zone height & \si{m} &
$U_{g,O_2}$ & Superficial gas velocity (air) & \si{m.s^{-1}}\\

$k_La^{CO_2}$ & Volumetric transfer coefficient (CO$_2$) & \si{s^{-1}} &
$V_{\mathrm{sump}}=A_{\mathrm{sump}}h_{\mathrm{sump}}$ & Sump zone volume & \si{m^3}\\

$k_La^{O_2}$ & Volumetric transfer coefficient (air) & \si{s^{-1}} &
$\dot{V}_e$ & Evaporative loss & \si{m^3.s^{-1}}\\

$k^{\mathrm{atm}}_{CO_2}$ & Atmospheric CO$_2$ exchange & \si{s^{-1}} &
$Y^{\mathrm{alg}}_{CO_2}$ & CO$_2$ yield per algal mass & \si{g.g^{-1}}\\

$k^{\mathrm{atm}}_{O_2}$ & Atmospheric $O_2$ exchange & \si{s^{-1}} &
$Y^{\mathrm{alg}}_{O_2}$ & $O_2$ yield per algal mass & \si{g.g^{-1}}\\

$k^{\mathrm{pw}}_{CO_2}$ & Paddlewheel CO$_2$ exchange & \si{s^{-1}} &
$y_{CO_2}$ & CO$_2$ fraction in air & --\\

$k^{\mathrm{pw}}_{O_2}$ & Paddlewheel $O_2$ exchange & \si{s^{-1}} &
$y_G$ & Gas molar fraction (Henry's law) & --\\

$k^{\mathrm{strip}}_{CO_2\leftarrow O_2}$ & Cross stripping ($O_2\to CO_2$) & -- &
$y_{O_2}$ & Oxygen fraction in air & --\\

$k^{\mathrm{strip}}_{O_2\leftarrow CO_2}$ & Cross stripping ($CO_2\to O_2$) & -- &
$y_{\mathrm{pure}\,CO_2}$ & Fraction of pure CO$_2$ & --\\

$K_{1,\mathrm{ref}}$ & First dissociation at \SI{25}{\celsius} & \si{mol.L^{-1}} &
$\alpha_{CO_2}$ & $k_La$ scale (CO$_2$) & --\\

$K_{2,\mathrm{ref}}$ & Second dissociation at \SI{25}{\celsius} & \si{mol.L^{-1}} &
$\alpha_{O_2}$ & $k_La$ scale ($O_2$) & --\\

$K_{H}^{CO_2}$ & Henry constant for CO$_2$ & \si{mol.m^{-3}.atm^{-1}} &
$\beta_{CO_2}$ & $k_La$ exponent (CO$_2$) & --\\

$K_{H}^{O_2}$ & Henry constant for oxygen & \si{mol.m^{-3}.atm^{-1}} &
$\beta_{O_2}$ & $k_La$ exponent ($O_2$) & --\\

$K_{H,\mathrm{ref}}^G$ & Henry constant at reference temperature & \si{mol.m^{-3}.atm^{-1}} &
$\Delta H_{K_1}$ & Enthalpy (first dissociation) & \si{J.mol^{-1}}\\

$K_{W,\mathrm{ref}}$ & Water autoprotolysis at \SI{25}{\celsius} & \si{(mol.L^{-1})^2} &
$\Delta H_{K_2}$ & Enthalpy (second dissociation) & \si{J.mol^{-1}}\\

$M_{CO_2}$ & CO$_2$ molar mass & \si{g.mol^{-1}} &
$\Delta H_{K_W}$ & Enthalpy (autoprotolysis) & \si{J.mol^{-1}}\\

\bottomrule
\end{tabularx}

\vspace{0.4\baselineskip}
\end{table*}

\endgroup

%\label{tab:params}
% \twocolumn
% --------------------------------------------------------

%%%%%%%%%%%%%%%%%%%%%%%%%%%%%%%%%%%%%%%%%%%%%%%%%%%%
% SECTION 3: BENCHMARK DESCRIPTION
%%%%%%%%%%%%%%%%%%%%%%%%%%%%%%%%%%%%%%%%%%%%%%%%%%%%
\section{Benchmark Description}
\label{sec:benchmark}
\raggedbottom

This section describes the operational structure of the proposed benchmark, detailing how users interact with the simulation environment and how controller performance is evaluated. First, the simulation loop that governs the execution of each experiment is presented, detailing the different steps that compose it. The required structure of user-implemented controllers is then outlined, specifying the interface, data flows, and constraints that ensure reproducibility across experiments. Finally, we define the cost functions and performance metrics computed by the benchmark to provide a standardized and quantitative assessment of control strategies. The benchmark is coded in Matlab programming language and available at GitHuB (https://github.com/guzmanjl/benchmarkmicroalgae).

\subsection{Simulation loop}\label{subsec:loop}

Users interact with the proposed benchmark through the script \texttt{Benchmark\_main.m}. The purpose of this script is to load the initial conditions and disturbance trajectories for the selected experiment, read the user-defined controller functions, execute the simulation, and present the resulting performance both graphically and through numerical indicators.

User interaction with the script consists of specifying the names of the functions that implement each of the control strategies. These controllers operate independently for each of the control tasks considered in the benchmark, namely pH regulation, dissolved oxygen control, simultaneous biomass concentration and water depth control, and temperature regulation. The mapping between manipulated variables and control objectives is predefined: CO$_2$ flow rate for pH control, air flow rate for dissolved oxygen, harvesting and dilution operation commands for biomass concentration and water depth, and heat exchanger flow rate together with inlet temperature for culture temperature. The required structure and interface of the controller functions are described in detail in Section \ref{subsec:controladores}.

Once the controller functions have been declared, the user can execute the script, which will automatically run the complete simulation. The simulation is performed by the function \texttt{simulate\_benchmark\_model.m}. This function concatenates the previously loaded data, defines the system constants, and then executes the simulation loop. The benchmark runs with a simulation step of $T_m=60$s. At each step $k$ the following information is available:

\begin{enumerate}
\item Outputs \((pH(k),DO(k),X_{alg}(k),Depth(k))\) are computed from the state.

\item The information about the current state of the system is packed in different structures depending on the nature of each variable: 
\begin{itemize}
    \item The structure $\mathrm{\mathbf{Timeline}}$ is generated with information related to the simulation time:
    \begin{equation*}
    \begin{aligned}
    \mathrm{\mathbf{Timeline}} = \{ & dt, index, time, \\
                           & time\_secday, hour, min \}.
    \end{aligned}
    \end{equation*}
    
    In the simulation loop, \emph{dt} denotes the numerical integration step, \emph{index} contains the current simulation step \emph{k}, \emph{time} represents the total elapsed time within the simulation, expressed in seconds, whereas \emph{time\_secday} provides the same quantity but relative to the current simulated day, cycling from 1 to 86400 before being reset. From this, the variables \emph{hour} and \emph{min} are derived, indicating the current simulated hour (0-24) and minute (0-60), respectively, both resetting at the end of their corresponding cycles.
    
    \item The structure $\mathrm{\mathbf{obs}}$ contains the present value of the main states of the system:
    \begin{equation*}
    \mathrm{\mathbf{obs}}=\{pH,DO,Depth,Xalg\_gL,T\}.
    \end{equation*}

    \item The structure $\mathrm{\mathbf{refs}}$ contains the fixed setpoints for the controlled variables: pH, DO, and culture temperature, set to 8, 150\%, and 30$^{\circ}$C, respectively, which are described for the default microalgae strain, {\it Scenedesmus almeriensis}, such as described in Section  \ref{subsec:reactor}.
    
    \item The structure $\mathrm{\mathbf{env}}$ contains the present value of the disturbances of the system:
    \begin{equation*}
    \begin{aligned}
    \mathrm{\mathbf{env}} = \{ & RadGlobal, RadPAR, \\
                      & Temp\_ext, RH, Wind \}.
    \end{aligned}
    \end{equation*}

    \item The structure $\mathrm{\mathbf{future}}$ groups all variables related to the preview information available to the controller, namely
    \begin{equation*}
    \begin{aligned}
    \mathrm{\mathbf{future}} = \{ & t\_future, RadGlobal, \\
                         & RadPAR, Temp\_ext, RH, Wind \}.
    \end{aligned}
    \end{equation*}
    In this case, the disturbance-related variables are represented as vectors containing their predicted future values from the current simulation time until the end of the horizon. The array \emph{t\_future} specifies the remaining time associated with each element of these vectors, thereby providing a consistent temporal reference for all forecasted disturbance profiles.
\end{itemize}

\item The four different user controllers follows the following structure:
\begin{equation*}
\begin{aligned}
[\mathrm{st\_CtrSignals}, \mathrm{st\_pH}] = \mathtt{controller\_pH}( ... & \\
\mathrm{Timeline}, \mathrm{obs}, \mathrm{refs}, \mathrm{env}, \mathrm{future}, ... & \\
\mathrm{st\_CtrSignals}, \mathrm{st\_pH}) & ,
\end{aligned}
\end{equation*}
which is defined for pH in this case. $\mathrm{\mathbf{st\_CtrSignals}}$ is a structure containing information about the 6 different control signals of the system, that will be detailed later along with $\mathrm{\mathbf{st\_pH}}$. This is performed similarly for each of the four controller functions, with the distinction that each of them has its own structure $\mathrm{\mathbf{st\_pH}}$, $\mathrm{\mathbf{st\_DO}}$, $\mathrm{\mathbf{st\_HV}}$ or $\mathrm{\mathbf{st\_T}}$, and function name $\mathtt{controller\_pH}$, $\mathtt{controller\_DO}$, $\mathtt{controller\_HV}$, or $\mathtt{controller\_T}$

\item The ouputs of the controllers are then saturated to satisfy the physical limits of the system, and gas commands \((Q_{\mathrm{air}},Q_{\mathrm{CO_2}})\) are delayed internally via an embedded First-Input First-Output (FIFO) buffer, and the plant uses the delayed signals \((Q_{\mathrm{air}}^{\mathrm{del}},Q_{\mathrm{CO_2}}^{\mathrm{del}})\) to simulate the time delay of the system.

\item The ODE is integrated over \([t,t+T_m]\) with constant inputs using a stiff ODE solver.
\end{enumerate}

After completing the simulation, the function computes the dynamic cost functions and the KPIs used to provide a quantitative assessment of the implemented controllers. These metrics summarize the deviation from setpoints, control effort, constraint violations, and overall system performance throughout the experiment. The entire process described above is fully automated to the user, whose sole responsibility is to design the controller functions according to the prescribed interface. All data handling, model integration, metric calculation, and result visualization are automatically managed by the benchmark framework.

\subsection{Controller Structure}\label{subsec:controladores}

To use the simulator correctly, it is essential to understand in detail how the controller functions are implemented. As an example, we consider the controller responsible for regulating pH through the injection of CO$_2$. As mentioned in Section~\ref{subsec:loop}, this function follows the structure shown below:

\begin{equation*}
\begin{aligned}
[\mathrm{st\_CtrSignals}, \mathrm{st\_pH}] = \mathtt{controller\_pH}( ... & \\
\mathrm{Timeline}, \mathrm{obs}, \mathrm{refs}, \mathrm{env}, \mathrm{future}, ... & \\
\mathrm{st\_CtrSignals}, \mathrm{st\_pH}) & ,
\end{aligned}
\end{equation*}

The user has access to information regarding the simulation time, the current values of all state variables, and both the present and future disturbance trajectories. In addition, the controller receives the structure $\mathrm{\mathbf{st\_pH}}$, which is initially empty. Its purpose is to allow the user to store any internal variables or auxiliary data required for the pH control strategy. Since this structure is returned as an output of the controller, it can be updated at each iteration of the simulation loop and subsequently read during the next controller call. Each controller is associated with its own dedicated state structure, which can be used to preserve any information the designer considers necessary.

To illustrate the use of this internal state variable, two examples are provided. The first and simplest one is a standard PI controller. Such a controller computes the CO$_2$ flow rate based not only on the instantaneous pH error, but also on the integral of that error. Since the current value of the integral term depends on its previous value, it is natural to store this variable inside ($\mathrm{\mathbf{st\_pH}}$):

\begin{verbatim}
if isfield(st_pH, interr)
    interr = st_pH.interr;
else
    interr = 0;
end

err = env.refs.pH - obs.pH;
interr = interr + Ki*err*dt;
st_pH.interr = interr;
st_CtrSignals.Qco2 = Kp*err + interr;
\end{verbatim}

The internal state structure can also be used to store past values of any input, since at each iteration the controller function only has direct access to the current states and disturbances. This capability is particularly useful for computing derivatives, when derivative action is desired, or for tasks such as system identification or the implementation of autoregressive models. For example, the following code stores and uses the previous pH measurement to compute its increment:

\begin{verbatim}
if isfield(st_pH, pH_k1)
    pH_k1 = st_pH.pH_k1;
else
    pH_k1 = obs.pH;
end

delta_pH = obs.pH - pH_k1;
st_pH.pH_k1 = obs.pH;
\end{verbatim}

Finally, the main output of each controller is the structure $\mathrm{\mathbf{st\_CtrSignals}}$, which is shared across all controllers. This structure contains six primary fields: $Qco2$, $Qair$, $Qd\_cmd$, $Qh\_cmd$, $Qhx$ y $Tin\_hx$. Note that the dilution and harvesting flows are binary variables, as they correspond to On/Off actuators. Every controller receives this structure both as an input and produces an updated version as output.

This design implies that controllers are not restricted to modifying only their `main' manipulated variable. For instance, if one wishes to implement a multivariable controller for pH and DO using the CO$_2$ and air flow rates, the implementation may reside entirely within the pH controller function: this function would write to $\mathrm{\mathbf{st\_CtrSignals}.Qco2}$ and $\mathrm{\mathbf{st\_CtrSignals}.Qair}$, while the DO controller function would simply refrain from modifying those fields. In this sense, the user has complete freedom to distribute control responsibilities across the four controller functions, provided that each manipulated variable is assigned a value at least once per simulation loop.

Once the desired controllers have been implemented, the user only needs to declare the name of the function containing each controller within the script \texttt{Benchmark\_main.m}. This automatically constructs the $\mathrm{\mathbf{Ctrl}}$ structure with the four corresponding function handles. If the user prefers to focus exclusively on a single control problem, the benchmark includes default controllers for the remaining tasks to ensure safe and nominal system operation. These default functions are \texttt{controller\_pH\_OnOff.m} for pH control, \texttt{controller\_DO\_OnOff.m} for DO control, \texttt{controller\_HD\_fixed.m} for harvesting and dilution, and \texttt{controller\_Temp\_HX\_no\_control.m} for temperature regulation.

All controller outputs must be expressed in m$^3$s$^{-1}$. For reference:
20 L min$^{-1}$ = (20/1000/60) m$^3$s$^{-1}$,
500 Lmin$^{-1}$ = (500/1000/60) m$^3$s$^{-1}$.

\subsection{Results evaluation}
\label{sec:evaluation}
This section describes the methodology used by the simulator to evaluate the performance of the controllers, both at the level of individual control loops and in terms of their overall impact on biomass production, productivity, and the consumption of $\mathrm{CO_2}$ and air.

First, dynamic cost functions associated with the control of pH, DO and temperature are defined. These functions quantify the performance of each control loop by providing a dimensionless, weighted metric that combines setpoint-tracking error, the smoothness of the control signal, and the use of resources (actuator flow rates). After introducing the cost functions, the KPIs are defined to quantify the cumulative consumption of $\mathrm{CO_2}$ and air, the productivity and biomass harvested metrics.

Such as commented above, the simulator assumes the following nominal operating conditions as optimal for biomass growth:
\[
\mathrm{pH}^\star = 8.0, \quad \mathrm{DO}^\star \leq 150~[\%\_{\mathrm{sat}}], \quad \mathrm{T}^\star = 30~[^\circ \mathrm{C}]
\]

\subsubsection{Dynamic Cost Functions}\label{subsec:DCF}
For each of the controlled variables, a dynamic cost function is defined that combines three partial objectives: setpoint-tracking error, the smoothness of the control signal, and the use of resources. In the following equations, $y(t)$ is the process variable, $y_{ref}(t)$ is the reference, and $u(t)$ is the control signal, for the corresponding control loop, pH, DO or T:

\begin{enumerate}
    \item \textbf{Setpoint tracking:}
    
    This is calculated by taking the cumulative absolute relative error of the variable to be controlled with respect to the setpoints, in order to penalise deviations from the optimal growth conditions.

    \begin{equation}
        J_{SP} = \sum_{k=1}^N{\left(\dfrac{|y_{ref}(k)-y(k)|}{y_{ref}(k)}\right)}
    \end{equation}
    
\item \textbf{Control-signal smoothness}:

The smoothness of the control signal is evaluated by penalizing large variations of the manipulated variable between consecutive sampling instants. The metric is defined as the sum of squared, normalized control increments with respect to the actuator operating range:
\begin{equation}
    J_{S} = \sum_{k=1}^{N} \left( \frac{u(k)-u(k-1)}{u_{\max}-u_{\min}} \right)^{2}
\end{equation}
By normalizing the increments with the actuator span $(u_{\max}-u_{\min})$, this index becomes dimensionless and comparable across different scenarios. Lower values of $J_{\text{smooth}}$ correspond to smoother control actions, whereas higher values indicate faster or more abrupt changes in the actuation over the simulation horizon.
    
    \item \textbf{Resource consumption:}

    The resource-consumption index is computed as the accumulated sum of the control signal normalized with respect to the maximum admissible flow rate of the actuator. The metric is defined as:
    \begin{equation}
        J_{C} = \sum_{k=1}^{N} \frac{u(k)}{u_{\max}}.
    \end{equation}
    By normalizing with $u_{\max}$, this index becomes dimensionless and can be used to quantify and penalize excessive use of the available resources. In the present application, this term is specifically introduced to discourage unnecessary consumption of $\mathrm{CO_2}$ and air demanded by the controllers.
\end{enumerate}

Finally, once the partial cost terms have been computed, a weighted sum of all metrics is evaluated in order to obtain a single scalar performance index. This overall cost is then used to assess the quality of the implemented control strategy, simultaneously taking into account the setpoint tracking capability, the smoothness of the control signals, and the associated resource consumption.

For the pH and DO control loops, the overall cost is defined as:
\begin{equation}
    J_{z} = W_{SP,z} \cdot J_{SP,z} + W_{S,z} \cdot J_{S,z} + W_{C,z} \cdot J_{C,z},
\end{equation}
with $z =\{pH,DO\}$ and where $J_{SP,z}$ measures the tracking performance, $J_{S,z}$ quantifies the smoothness of the (single) control signal, and $J_{C,z}$ represents the resource consumption, for pH and DO, respectively.

In contrast, for the temperature control loop two manipulated variables are involved (the inlet temperature to the heat exchanger and the corresponding flow rate). In this case, the cost function penalizes both control signals through two contributions of the smoothness term, leading to:
\begin{equation}
    J_{Temp} = W_{SP,T} \cdot J_{SP,T} + W_{S1,T} \cdot J_{S,T} + W_{S2,T} \cdot J_{S,T} + W_{C,T} \cdot J_{C,T}
\end{equation}
The weighting coefficients $W_{SP,T}$, $W_{S,T}$, $W_{S1,T}$, $W_{S2,T}$ and $W_{C,T}$ are fixed by the designer, remain unknown to the user, and cannot be modified.

Finally, a global performance index $J_{\text{avg}}$ is computed as the average of the individual cost functions associated with the loops of pH ($J_{pH}$), DO ($J_{DO}$), and T ($J_{Temp}$), in order to obtain a single quantitative measure of the overall control performance.

\subsubsection{Key Performance Indicators (KPIs)}\label{subsec:KPI}
The simulator computes a set of KPIs to summarize the overall system behaviour in terms of biomass growth, harvesting performance and resource consumption. The following KPIs are reported:

\begin{itemize}
    \item \textbf{Total air injected} [L]:  
    Cumulative volume of air supplied to the reactor over the entire simulation horizon. This indicator reflects the total aeration effort required by the control strategy:
  \begin{equation}
    \text{Total air injected} = \sum_{k=1}^{N} Q_{\mathrm{air}}(k)\cdot T_m
    \end{equation}
    
    \item \textbf{Total CO$_2$ injected} [L]:  
    Cumulative volume of CO$_2$ injected into the reactor during the simulation. It quantifies the overall CO$_2$ consumption associated with pH and growth control:
  \begin{equation}
    \text{Total $\mathrm{CO_2}$ injected} = \sum_{k=1}^{N} Q_{\mathrm{CO_2}}(k)\cdot T_m
    \end{equation}
    
    \item \textbf{Harvested amount} [g]:  
    Total biomass removed from the system through harvesting operations during the simulation:
    \begin{equation}
    \text{Harvested} = \sum_{k=1}^{N} X_{alg}(k)\cdot Q_h(k)\cdot T_m
    \end{equation}
    
    \item \textbf{Total biomass produced} [g]:  
    Total amount of biomass generated during the simulation, combining the net biomass gain inside the reactor and the harvested biomass. It is computed as
    \begin{equation}
        \text{Biomass produced} =(X_{f} - X_{0}) + \text{Harvested}
    \end{equation}
    where $X_{0}$ and $X_{\mathrm{f}}$ denote the initial and final biomass in the reactor [g], and $\text{Harvested}$ is the cumulative harvested biomass [g].

    \item \textbf{Productivity per unit area} [g\,m$^{-2}$\,day$^{-1}$]:  
    Areal productivity based on the total biomass produced, normalized by the reactor area and the simulated time:
    \begin{equation}
        \text{Prod area} = \frac{\text{Biomass produced}}{A \  t_{\mathrm{days}}}
   \end{equation}
    where $A$ is the reactor area [m$^{2}$] and $t_{\mathrm{days}}$ is the simulated time in days.

    \item \textbf{Production yield ratio} [\%]:  
    Fraction of the biomass produced that is effectively harvested, expressed as a percentage:
    \begin{equation}
        \text{Yield} = 100 \cdot \frac{\text{Harvested}}{\text{Biomass produced}}
    \end{equation}
        
       Values close to 100\% indicate that almost all of the biomass produced is harvested, while lower values reveal significant accumulation within the reactor and higher values reveal that more biomass has been harvested than produced.

    \item \textbf{Harvested per unit area} [g\,m$^{-2}$\,day$^{-1}$]:  
    Areal productivity of the harvested biomass only, normalized by area and simulated time:
    \begin{equation}
        \text{Harv area} = \frac{\text{Harvested}}{A \, t_{\mathrm{days}}}
     \end{equation}

    \item \textbf{Relative biomass accumulation} [\% of initial]:  
    Relative net change of biomass inside the reactor with respect to its initial value:
     \begin{equation}
        \text{Accum rel} = 100 \cdot \frac{X_{\mathrm{f}} - X_{0}}{X_{0}}.
     \end{equation}
    Positive values indicate net accumulation of biomass in the reactor, whereas negative values (e.g., \mbox{$-2.57$\,\%}) correspond to a net decrease with respect to the initial biomass.
\end{itemize}

\subsection{Simulation output structure}
\label{subsec:sim-output}

The benchmark simulator returns its results in a structured MATLAB variable
\texttt{results}, which groups time series, control actions, performance
indicators, and additional internal states needed for analysis and
post-processing. The main fields and meanings can be found in \ref{app:results-structure}.

This structured output allows direct comparison across different control
strategies, straightforward computation of benchmark indices, and efficient
post-processing of the simulations within MATLAB or other data-analysis
environments.

% \subsection{Embedded Gas-Transport Delay}
% Air and CO\(_2\) injections are delayed by a conveyor (FIFO) buffer of length
% \[
% n_{\mathrm{delay}}=\mathrm{round}\!\left(\frac{L/v_{\mathrm{liq}}}{T_m}\right),
% \]
% using liquid path \(L\) and velocity \(v_{\mathrm{liq}}\). The delay is fully internal to the simulator (users do not configure it).

% \subsection{Reproducibility Notes}
% \begin{itemize}
% \item \emph{Fixed-step integration.} The ODE is integrated over each step with constant inputs via a stiff BDF method.
% \end{itemize}

%% Controllers Under Test
%%\section{Controllers Under Test}
%%\label{sec4}

%%\paragraph{Controller classes.} 
%%\emph{Bang–bang:} hysteresis rules on pH and DO with anti-windup timers; 
%%\emph{PID:} independent PIDs on pH and DO with decoupling/feedforward on $PAR$; 
%%\emph{MPC:} linear or nonlinear MPC on $(\mathrm{pH},DO)$ with input and rate constraints. The benchmark package provides identical initial conditions and parameter sets to ensure reproducibility and comparability across designs \citep{Melo2022_open_benchmarks,Montesuma2023_tep_benchmark}. 

%% Results and Discussion
\section{Results}
\label{sec5}

This section presents four benchmarking scenarios for four different players using alternative control algorithms. The simulation results are individually described and quantitatively compared using the metrics described in Section \ref{sec:evaluation}, which are summarized in Table~\ref{tab:indices}. The code for running the first player is available in GitHub for a better understanding. 

\subsection{First player: classical On/Off operation}
\label{subsec:first_player}
\begin{figure}
    \centering
    \includegraphics[width=\linewidth]{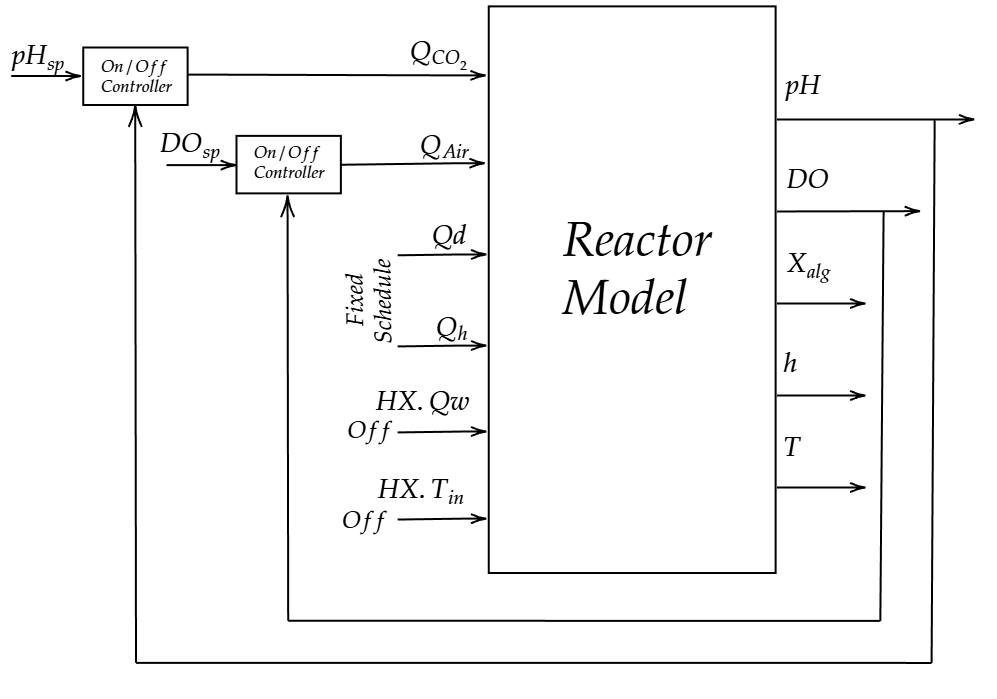}
    \caption{Control scheme used in Player 1 configuration}
    \label{fig:1P_scheme}
\end{figure}

The first player represents the baseline operation commonly used in open raceway reactors, relying exclusively on simple On/Off logic for process regulation and a fixed harvesting-dilution schedule as shown in the diagram in Figure \ref{fig:1P_scheme}. This configuration does not employ model-based tuning or dynamic adaptation, and therefore serves as a reference against which the more advanced players can be evaluated.

In this strategy, pH is regulated by activating CO$_2$ injection whenever the measured value exceeds a fixed upper threshold, and deactivating it when the pH returns below a lower bound. This hysteresis band avoids excessive switching but results in the characteristic saw-tooth pattern typically observed in industrial raceways, where pH rises steadily during daylight due to photosynthetic alkalinity uptake and is periodically corrected by short CO$_2$ pulses.

\begin{figure*}
    \centering
    \includegraphics[width=\linewidth]{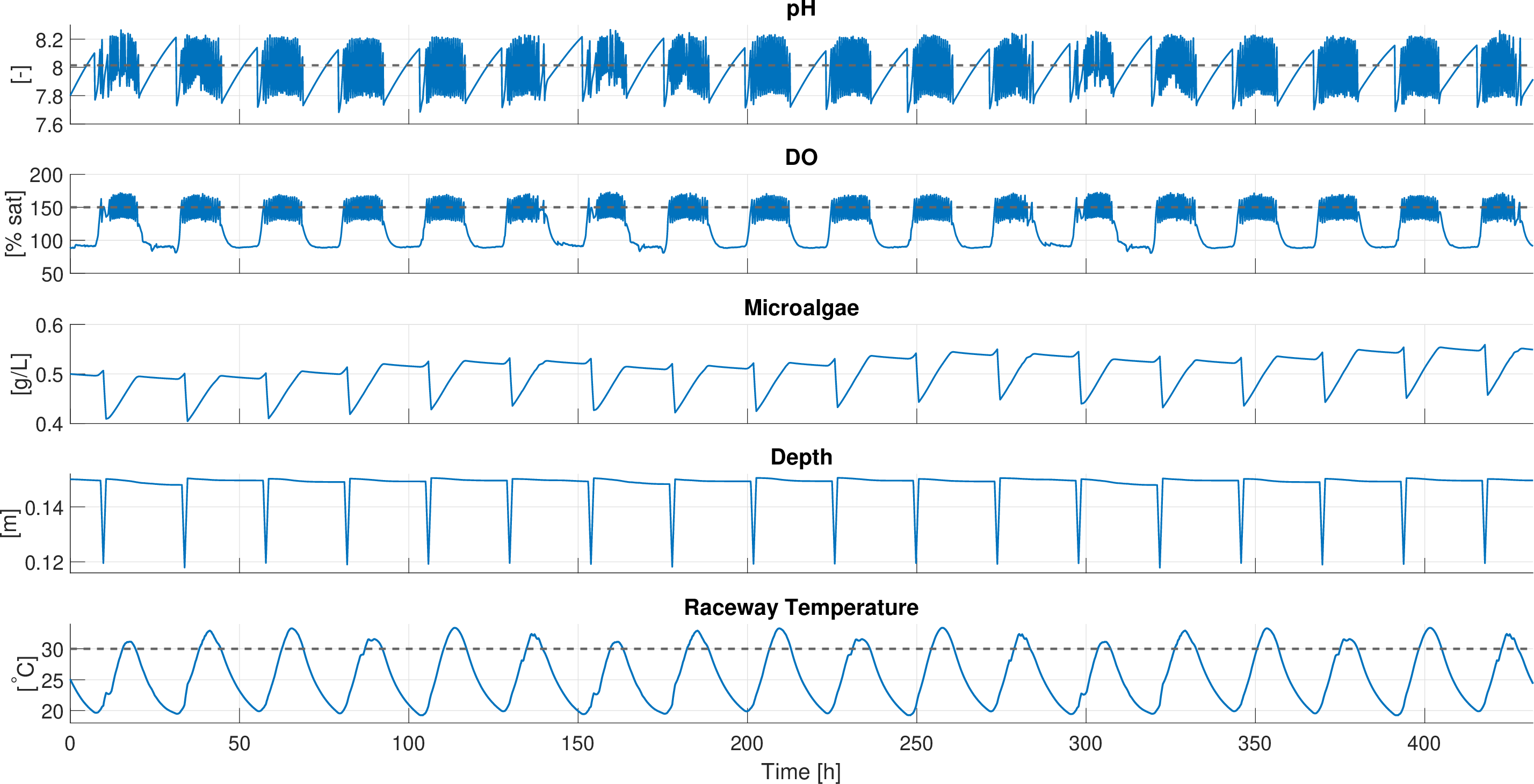}
    \caption{Player 1 simulated trajectories. From top to bottom: pH, DO, biomass concentration, water depth and water temperature trajectories. In blue, the variable value. In dashed black, the setpoints.}
    \label{fig:trajectories-player1}
\end{figure*}

DO is controlled in a similar manner. When the DO concentration surpasses a predefined limit (150\% saturation), aeration is switched on to enhance gas-liquid mass transfer and relieve supersaturation. Once DO falls below the threshold, aeration is turned off again. This loop acts primarily as a protective mechanism to avoid extreme DO values, rather than as a controller designed to track a specific reference.

The harvesting and dilution strategy follows a fixed daily procedure. At 09:00, the system enters a harvesting phase that removes culture until the water depth has decreased by a dilution rate of 20\%. Immediately afterwards, a dilution phase restores the nominal depth by adding fresh medium. Both actuators operate in binary mode, and the same volumes are processed each day regardless of the environmental conditions or culture state.

No temperature regulation is included in this player. The heat exchanger remains inactive at all times, so the reactor temperature evolves freely according to the environmental heat balance. As a consequence, significant fluctuations can occur depending on weather conditions.

Overall, this first player provides a clear representation of classical raceway operation, with minimal sensing requirements and simple hysteresis-based actuation. Its performance is used as the baseline for comparison with more advanced control approaches. The trajectories obtained for this player, together with the associated key performance indicators, are presented in Figure~\ref{fig:trajectories-player1} and Table~\ref{tab:indices}. Notice that the dynamic cost functions in the table are normalized with respect to this player for a better comparison with other strategies used on the system. The KPIs reported characterize the performance of this baseline strategy in terms of productivity and operability. They summarize the overall air and CO$_2$ consumption, the total biomass produced, the productivity per unit area (which is particularly useful for comparing different control strategies), and the total amount of biomass harvested over the simulation horizon. In addition, the production yield ratio indicates whether the harvested biomass has been effectively generated during the simulated period or rather extracted from the biomass present in the initial conditions, while the relative biomass accumulation quantifies the fraction of the initial biomass that remains in the reactor at the end of the simulation.

\subsection{Second player: PI-based regulation with thermal On/Off control}
\label{subsec:second_player}
\begin{figure}
    \centering
    \includegraphics[width=\linewidth]{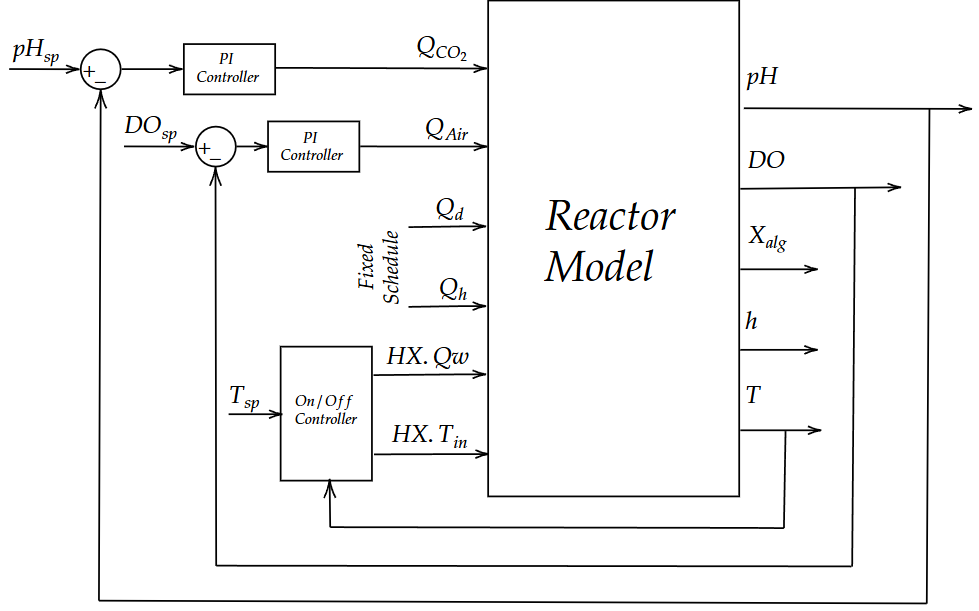}
    \caption{Control scheme used in Player 2 configuration}
    \label{fig:2P_scheme}
\end{figure}

The second player builds directly upon the baseline configuration by introducing proportional-integral (PI) control for pH and DO, while maintaining the same harvesting and dilution strategy used for the first player (see Figure \ref{fig:2P_scheme}). In addition, a simple thermal regulation mechanism is incorporated through an On/Off actuation of the heat exchanger. The objective of this player is to assess how basic feedback control, supported by process-model identification, improves process stability and overall performance relative to classical On/Off operation.

%To implement the PI loops, first-order-plus-dead-time (FOPDT) models are identified from the behaviour of the first player, capturing the dynamic response of pH and DO to CO$_2$ injection and aeration, respectively.\textcolor{red}{The pH process was characterized by a steady-state gain of $K_{pH} = -2798.7$, a time constant of $\tau_{pH} = 1051~\mathrm{s}$, and a transport delay of $\theta_{pH} = 400~\mathrm{s}$. Likewise, the dissolved oxygen dynamics were approximated by a first-order model with gain $K_{DO} = -24460$, time constant $\tau_{DO} = 3773.6~\mathrm{s}$, and dead time $\theta_{DO} = 400~\mathrm{s}$.}
%\textcolor{red}{Based on these identified models, independent PI controllers were designed for each control loop using the SIMC tuning method. For the pH control loop, the controller parameters were set to a proportional gain of $K_{c,pH} = -6.15\times 10^{-4}$ and an integral time of $T_{i,pH} = 1051~\mathrm{s}$. Similarly, the dissolved oxygen controller was tuned with $K_{c,DO} = -1.98\times 10^{-4}$ and an integral time of $T_{i,DO} = 3109~\mathrm{s}$. The SIMC-based tuning ensures a robust closed-loop performance, providing a good compromise between disturbance rejection and robustness to model uncertainty \citep{skogestad2003}.}

To implement the PI loops, first-order-plus-dead-time (FOPDT) models are identified from the behaviour of the first player. Since both pH and dissolved oxygen (DO) are influenced by solar radiation as well as by the injection of CO$_2$ and air, respectively, two dynamic models are identified for each variable: one describing the effect of solar radiation and another capturing the response to gas injection. For control design purposes, the focus is placed on the models associated with gas injection, as these correspond to the manipulated variables of the pH and DO control loops.

Accordingly, the pH process related to CO$_2$ injection was characterized by a steady-state gain of $K_{pH} = -2798.7~\mathrm{s~m^{-3}}$, a time constant of $\tau_{pH} = 1051~\mathrm{s}$, and a transport delay of $\theta_{pH} = 400~\mathrm{s}$. Likewise, the dissolved oxygen dynamics associated with aeration were approximated by a first-order model with gain $K_{DO} = -24460~\mathrm{\%~s~m^{-3}}$, time constant $\tau_{DO} = 3773.6~\mathrm{s}$, and dead time $\theta_{DO} = 400~\mathrm{s}$.

Based on these identified models, independent PI controllers were designed for each control loop using the SIMC tuning method \citep{skogestad2003}. For the pH control loop, the controller parameters were set to a proportional gain of $K_{c,pH} = -6.15 \times 10^{-4}~\mathrm{m^3 s^{-1}}$ and an integral time of $T_{i,pH} = 1051~\mathrm{s}$. Similarly, the dissolved oxygen controller was tuned with $K_{c,DO} = -1.98 \times 10^{-4} ~\mathrm{m^3~s^{-1}~\%^{-1}}$ and an integral time of $T_{i,DO} = 3109~\mathrm{s}$.

%These models provide the characteristic time constant, process gain, and delay necessary to apply the SIMC tuning rules \citep{skogestad2003}, resulting in controller parameters that balance disturbance rejection and tracking performance without requiring extensive manual adjustment.

For pH regulation, the PI controller drives the CO$_2$ injection flow to counteract the strong diurnal alkalinity variations produced by photosynthesis. Compared to the purely hysteresis-based operation of the first player, this approach significantly reduces the amplitude of pH oscillations and prevents the large morning excursions typically observed under high irradiance.

DO regulation follows the same methodology. The identified model reflects the combined effects of photosynthetic oxygen production, biomass respiration, and gas–liquid mass transfer. The resulting PI controller modulates the aeration flow to maintain DO closer to its target value (150\% saturation), avoiding the abrupt transitions and overshoots inherent to On/Off control and providing smoother trajectories throughout the day.

The harvesting and dilution strategy remains identical to that of the first player. At 09:00, the system performs a harvesting phase until culture depth decreases by 20\%, followed by a dilution phase that restores the nominal water level. Keeping this component unchanged ensures that the performance differences are attributable solely to the upgraded feedback controllers.

A thermal regulation mechanism is also added in this player. Temperature is controlled by switching between a hot water stream at 50,$^{\circ}$C, supplied by a solar-heated storage tank, and a cold stream at 20,$^{\circ}$C from the water network. The controller activates the cold stream when the reactor temperature exceeds an upper threshold, and the hot stream when it falls below a lower threshold. Although the logic is binary, this addition prevents excessive thermal fluctuations and maintains the culture within a more favourable operating range.

\begin{figure*}
    \centering
    \includegraphics[width=\linewidth]{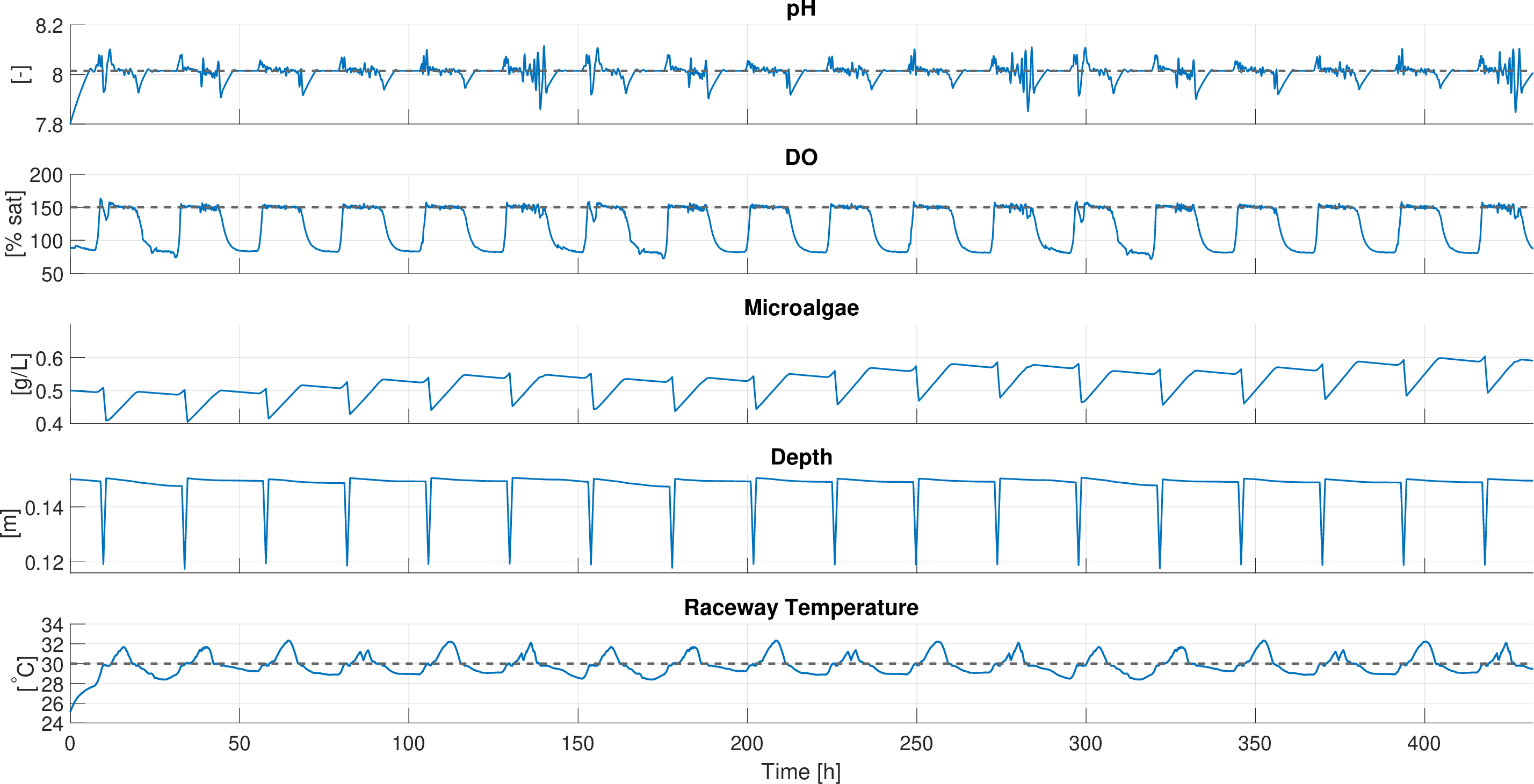}
    \caption{Player 2 simulated trajectories. From top to bottom: pH, DO, biomass concentration, water depth and water temperature trajectories. In blue, the variable value.  In dashed black, the setpoints.}
    \label{fig:trajectories-player2}
\end{figure*}

The trajectories obtained with this player highlight the impact of closing the loop around the main physico-chemical variables. Both pH and DO exhibit substantially smoother profiles, with reduced overshoots and improved stability, while temperature remains closer to its desired band as can be seen in Figure~\ref{fig:trajectories-player2}.

These improvements are reflected in the dynamic cost functions: the indices associated with pH and DO regulation decrease markedly with respect to Player~1, evidencing the positive impact of introducing PI control for these variables together with the On/Off temperature regulation, resulting in $J_{\text{avg}} = 0.4648$.  The KPIs in Table~\ref{tab:indices} show that this strategy achieves higher productivity per unit area and, consequently, an increase in harvested biomass of approximately 1 kg during the simulation horizon compared to the reference case

\subsection{Third player: constant biomass concentration}
\label{subsec:third_player}
\begin{figure}
    \centering
    \includegraphics[width=\linewidth]{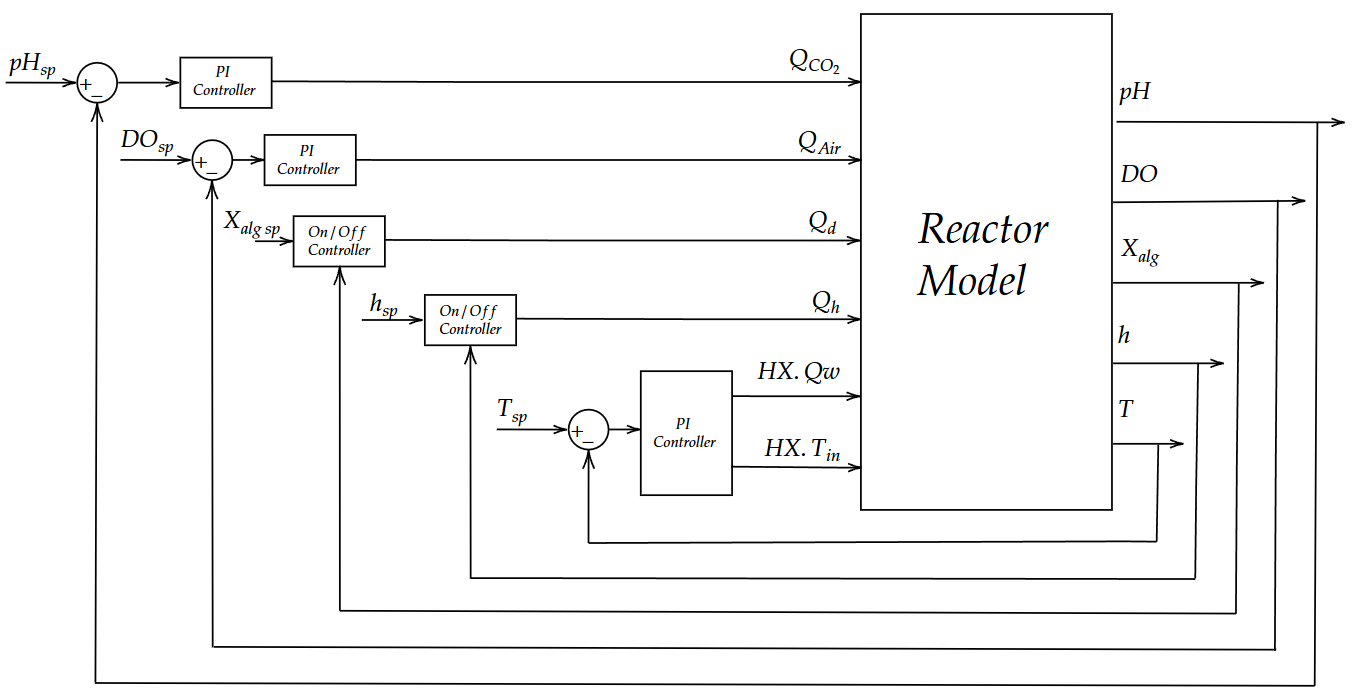}
    \caption{Control scheme used in Player 3 configuration}
    \label{fig:3P_scheme}
\end{figure}

This player is a more advanced version of the second player, including the same controllers for pH, and DO, but employing a different approach to harvesting and dilution, and implementing a PI controller for temperature regulation. In this case, instead of running a daily open-loop cycle that always harvests the same volume, regardless of the weather conditions on that specific day, two closed-loop strategies are established, with the intention of continuously regulating water depth and biomass concentration (see Figure \ref{fig:3P_scheme}). This method is commonly referred to as operation in turbidostat mode, adopted from bacterial cultivation, where the system is controlled to maintain an approximately constant turbidity, which is used as a direct measure of biomass concentration.

Given the binary character of these actuators, these strategies will be On/Off controllers. The variables are paired according to the most direct effect they have. In this case, the dilution flow rate is the only one of the two variables that has a direct effect on the biomass concentration, so both variables will be paired. In contrast, the harvesting flow rate only affects the water depth, so both variables will form the second pair.

For temperature regulation, the third player replaces the On/Off thermal logic of the second player with a fully continuous PI controller acting on the heat exchanger. The controller is designed using a FOPDT model identified from a step test in which the heat-exchanger water flow rate is perturbed while maintaining the inlet temperature at $50^{\circ}$C. The identified temperature model is characterized by a steady-state gain of $K_{T} = 19600~\mathrm{^{\circ}C~s~m^{-3}}$, a time constant of $\tau_{T} = 2850~\mathrm{s}$, and a transport delay of $\theta_{T} = 468~\mathrm{s}$. Following the same SIMC-based tuning methodology adopted for the pH and DO control loops, the temperature PI controller parameters were set to a proportional gain of $K_{c,T} = 1.4\times 10^{-4} ~\mathrm{m^3~s^{-1}~^{\circ}C^{-1}}$ and an integral time of $T_{i,T} = 2850~\mathrm{s}$. This tuning ensures a robust and smooth closed-loop response, providing effective temperature tracking and disturbance rejection while preventing excessive actuation of the heat-exchanger system.

\begin{figure*}
    \centering
    \includegraphics[width=\linewidth]{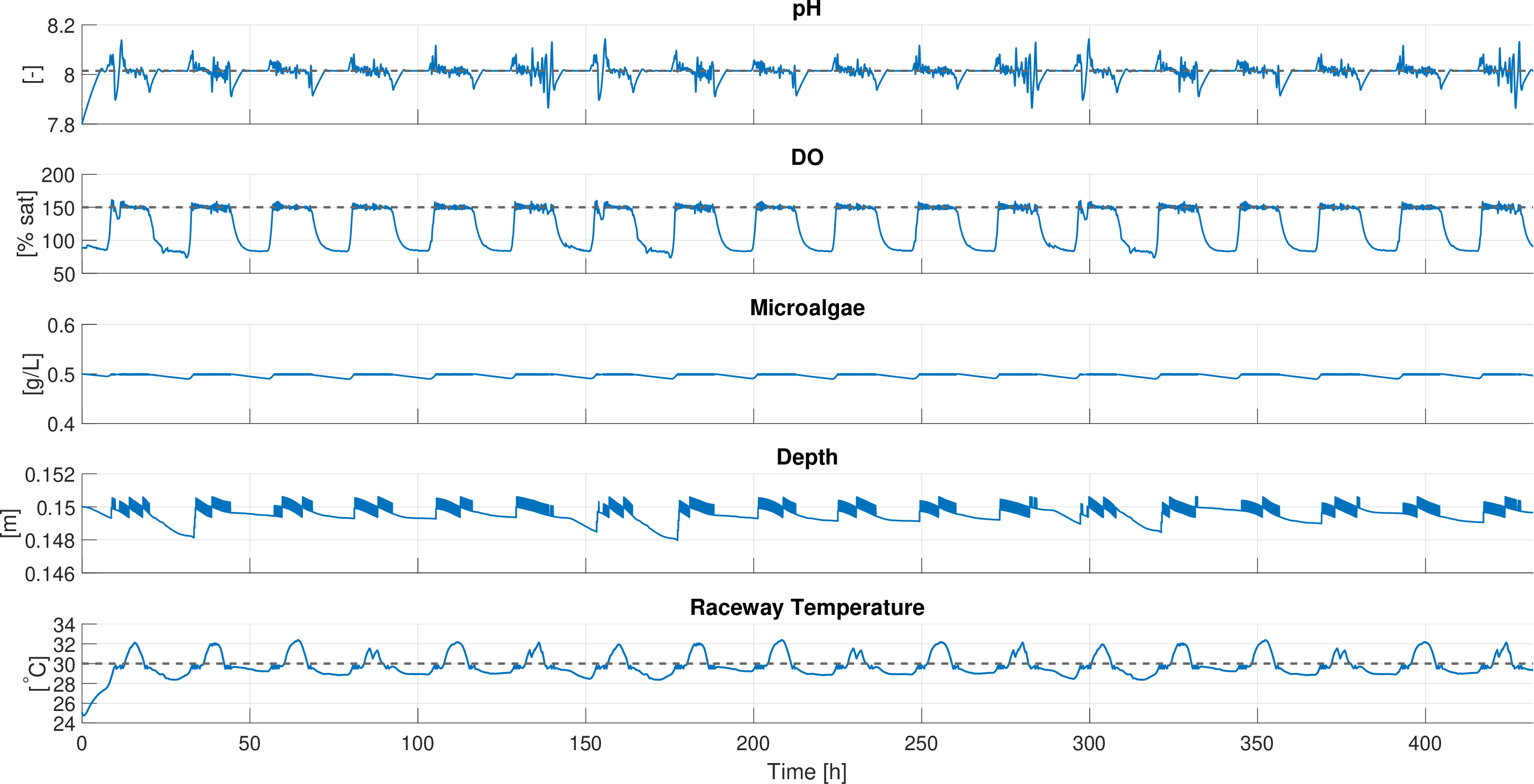}
    \caption{Player 3 simulated trajectories. From top to bottom: pH, DO, biomass concentration, water depth and water temperature trajectories. In blue, the variable value.  In dashed black, the setpoints.}
    \label{fig:trajectories-player3}
\end{figure*}

As references, both variables will be kept at a constant level, equal to their initial value. This allows for sustainable operation, regardless of environmental conditions. To this end, the first controller will activate the dilution actuator when the biomass concentration exceeds its initial value (0.5 g L$^{-1}$), and deactivate it when it drops below this value. A similar operation will be performed with the harvesting actuator, which will be activated when the water depth exceeds the initial value (15 cm). The results of this simulation are presented in Figure \ref{fig:trajectories-player3}.

The results in the Table~\ref{tab:indices} show pH and DO results with dynamic cost functions very similar to those of Player~2, as the same PI controllers are used in both cases. However, the performance achieved in temperature dynamic cost function is markedly superior to that obtained with the previous On/Off strategy, leading to a lower average global cost, with $J_{\text{avg}} = 0.3905$. In the context of KPIs, it is worth noting that the production yield index is slightly above 100\%, indicating that not much more biomass is being harvested than is being produced. This metric correlates with the relative biomass accumulation, which is also close to 0\%. All of this is indicative of sustainable operation that does not compromise the long-term productivity of the reactor.

On the other hand, productivity per unit area is higher than in previous strategies, despite having the same control over the other variables. This is because, in general, under sunny conditions similar to those in the proposed scenario, higher biomass concentrations provide higher productivity. Therefore, avoiding the sudden drop that occurs at the beginning of the day in the other players is very beneficial for the productivity of the system.

Following this same logic, this same strategy has also been evaluated with a higher biomass concentration reference. Increasing this reference to 0.8 g L$^{-1}$ provided a productivity per unit area of 21.92 g m$^{-2}$ day, higher than the previous one. This reinforces the importance of controlling this variable, even leading to the implementation of optimization strategies.

%The resulting PI controller provides smoother and more effective actuation of the heat-exchanger flow, reducing the strong thermal cycling observed under On/Off operation and maintaining the reactor temperature closer to the desired reference throughout the day.

\subsection{Fourth player: EMPC biomass optimization}
\label{subsec:fourth_player}
The fourth player introduced in this work incorporates optimization algorithms into the simulation framework. While this player does not add new control methods for pH, DO, or temperature, it implements a dynamic optimization strategy for biomass concentration based on Economic Model Predictive Control (EMPC). The objective of this strategy is to determine the optimal dilution profiles to maximize harvested biomass. For this purpose, the optimizer uses a simplified version of the process model, retaining only the biomass concentration as a dynamic state and fixing all other process variables to their nominal setpoints: pH 8, DO 150\%, water depth 15 cm, and temperature 30$^o$C. This simplification reduces the computational burden of the optimization problem.

\begin{figure}
    \centering
    \includegraphics[width=\linewidth]{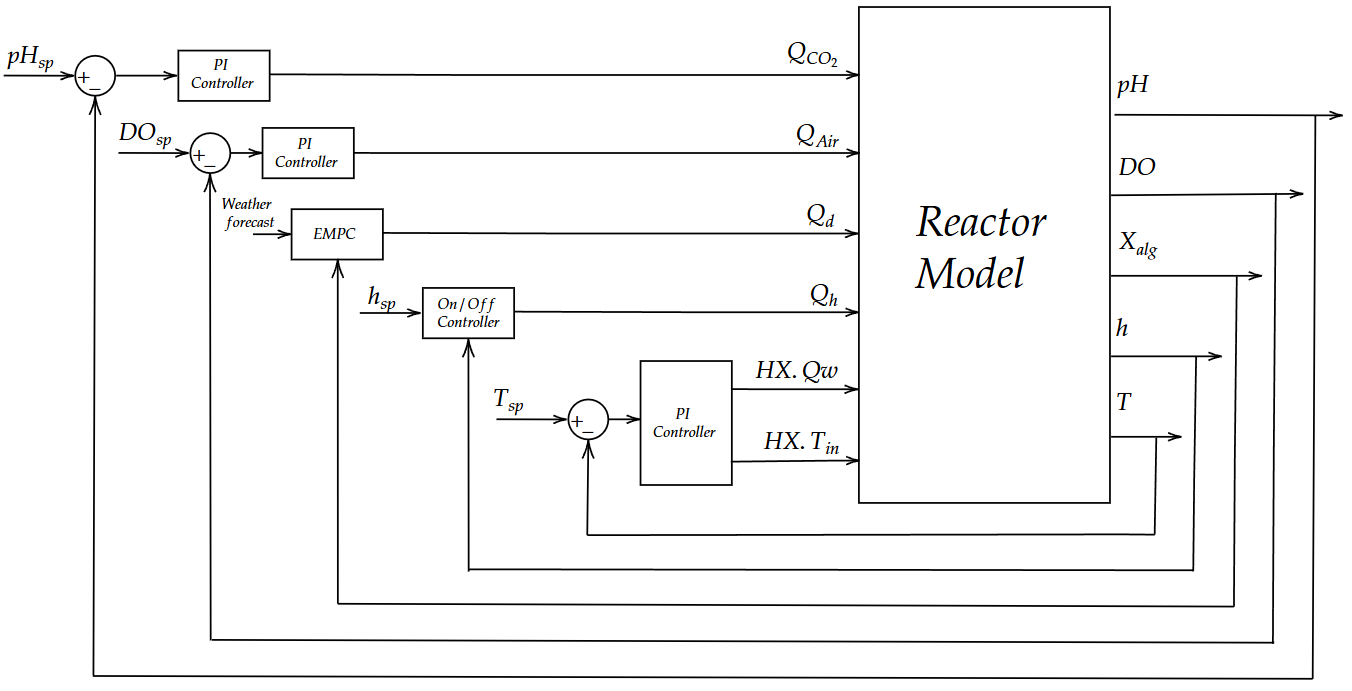}
    \caption{Control scheme used in Player 4 configuration}
    \label{fig:4P_scheme}
\end{figure}

\begin{figure*}
    \centering
    \includegraphics[width=\linewidth]{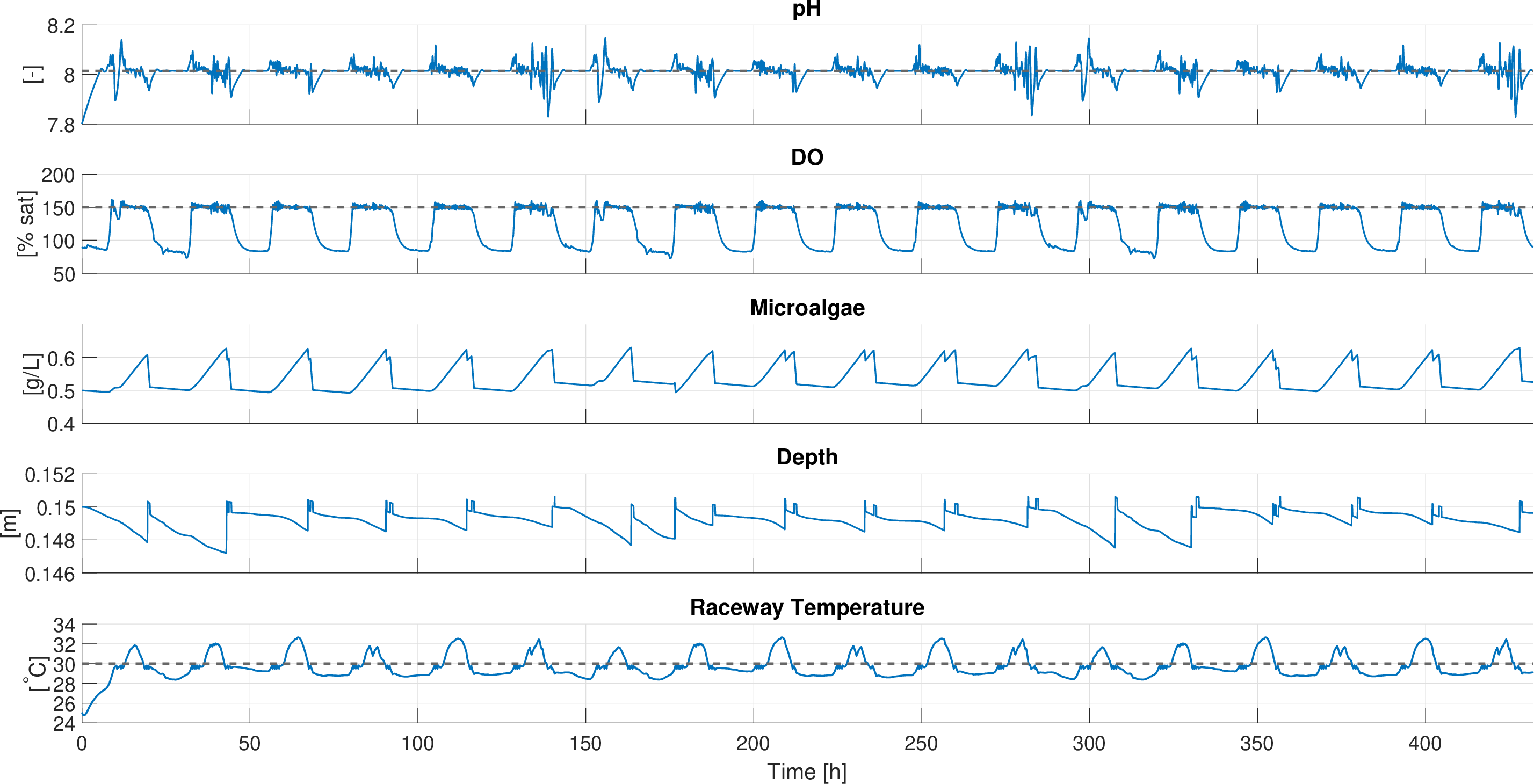}
    \caption{Player 4 simulated trajectories. From top to bottom: pH, DO, biomass concentration, water depth and water temperature trajectories. In blue, the variable value.  In dashed black, the setpoints.}
    \label{fig:trajectories-player4}
\end{figure*}

A detailed discussion of EMPC theory is beyond the scope of this paper and users are referred to \cite{Otalora2024Modelingcontrol} for further information. It suffices to note that the optimization problem is solved every 15 minutes using a prediction horizon, $N$, of one hour, i.e. four samples, and weather forecast is considered in the optimization algorithm (see the control scheme in Figure \ref{fig:4P_scheme}). The resulting optimal sequence consists of four binary decisions: whether to dilute at each 15-minute interval. Only the first action of the optimal sequence is implemented, and the problem is re-solved 15 minutes later with a receding horizon approach. This procedure is activated daily once solar radiation first exceeds 100 W m$^{-2}$, as optimization is only meaningful during periods of active biomass growth, and it is deactivated once radiation drops below this threshold for the last time that day. As the end of the day approaches, the remaining horizon naturally shrinks until only a single sample remains.

The cost function used by the optimizer is:

\begin{equation}
    J=\sum_{k=1}^N-P_{alg}\cdot X_{alg}(k)\cdot Q_d(k)
\end{equation}
where $P_{alg}$ is the selling price per gram of harvested biomass. The dilution flow $Q_d$ is used instead of the harvesting flow $Q_h$ because, under constant-level operation, both flows are practically identical, with evaporation losses assumed negligible.

The optimization problem includes a constraint on the manipulated variable: its value must lie within [0,1] since it represents a binary dilution decision. Additionally, a terminal constraint requires the biomass concentration at the end of the prediction horizon to remain above 0.5 g L$^{-1}$ . This constraint ensures sustainable reactor operation; without it, the optimizer would always prefer to dilute as much as possible, as this maximizes the economic objective.

The harvesting flow rate continues to be governed by water-depth control, following the same On/Off logic used for the previous player, with a threshold at 15 cm. The simulation results for this strategy are shown in Fig. \ref{fig:trajectories-player4}.

As can be observed, the optimizer tends to concentrate dilution actions near the end of the day. This behavior arises because, at this time of year, higher biomass concentrations are beneficial during midday when radiation is strongest. Toward the end of the day, however, the optimizer systematically harvests down to the 0.5 g L$^{-1}$ limit to maximize economic performance. According to the Table \ref{tab:indices}, the dynamic cost functions obtained for this strategy are very similar to those of Player~3, with slightly higher values, $J_{\text{avg}} = 0.4047$. This is expected, since Player~4 is primarily designed to maximize production: during EMPC operation, the biomass concentration departs from the nominal value used to tune the PI controllers. This mismatch highlights the need for adaptive models capable of maintaining control performance over a wider range of biomass concentrations.
However, the KPIs indicated that this strategy achieves the highest productivity and biomass harvested per unit area, as well as the highest total biomass harvested, 4 kg higher than player 1 and almost 2 kg higher than player 3, while maintaining a positive cumulative biomass balance.

It is important to note that this strategy does not surpass the productivity achieved by Player 3 when the biomass concentration setpoint is increased to 0.8 g L$^{-1}$. This limitation is primarily due to the terminal constraint requiring the biomass concentration to remain above 0.5 g L$^{-1}$ . Nevertheless, alternative optimization approaches could be explored, such as modifying the cost function to reward biomass growth rather than harvested biomass, or employing longer prediction horizons that might yield greater medium-term benefits.

%% Conclusions
\section{Conclusions}
\label{sec6}

This work has presented a new contribution to the process control community by introducing a high-fidelity benchmark platform specifically designed for Outdoor Microalgae Raceway Reactors (ORPs). This platform addresses a critical gap by providing an openly available system that simultaneously captures four coupled regulation tasks: pH, dissolved oxygen (DO), thermal control via a heat exchanger, and hydraulic management via coordinated harvest-dilution actions, under realistic multi-day outdoor disturbances. The integration of an experimentally calibrated dynamic model ensures that the benchmark accurately reflects the coupled thermal, physicochemical, and biological processes, including complex aspects like stiff dynamics, multirate disturbances, and actuator saturation, making it a demanding testbed for control strategies.

The evaluation of the four baseline regulatory architectures highlights the practical utility of the platform and provides clear insights into the benefits of advancing control complexity:

\begin{enumerate}
    \item Transition from On/Off to PI Control (Player 1 vs. Player 2): Implementing PI control for pH and DO, tuned via simple model-based rules, substantially improved stability compared to classical On/Off operation. The PI controller significantly reduced the amplitude of pH oscillations and prevented the large morning excursions characteristic of hysteresis-based control. This demonstrates that basic feedback mechanisms, supported by process identification, are necessary for achieving smoother and more stable control of the main physicochemical variables.

    \item Implications of Closed-Loop Biomass Management (Player 3): Moving away from fixed daily schedules (Players 1 and 2) to a closed-loop Turbidostat strategy (Player 3) demonstrated significant operational advantages. This strategy successfully maintained the culture at a constant biomass concentration, achieving a sustainable operation with a production yield ratio very close to $100\%$ and relative biomass accumulation near $0\%$. Crucially, avoiding the sudden biomass drops seen in the fixed-schedule players resulted in a higher areal productivity ($\text{17.64 g m}^{-2} \text{ day}^{-1}$). Further testing showed that increasing the biomass concentration setpoint to $0.8 \text{ g L}^{-1}$ further increased productivity to $\text{21.92 g m}^{-2} \text{ day}^{-1}$, reinforcing the importance of optimal biomass control.

    \item Potential of Optimization (Player 4): The implementation of the EMPC strategy (Player 4) demonstrated the potential for embedding optimization directly into the control architecture. By dynamically optimizing the dilution profile to maximize harvested biomass, this strategy achieved the highest total harvested biomass and the highest areal productivity ($\text{19.01 g m}^{-2} \text{ day}^{-1}$) among the baseline strategies constrained to maintaining a minimum $0.5 \text{ g L}^{-1}$ biomass level. The optimization logic favored concentrating dilution actions near the end of the day to maximize economic return while respecting the sustainability terminal constraint.
\end{enumerate}

Overall, the established evaluation framework, centered on the unified performance index $J$ and comprehensive Key Performance Indicators (KPIs) like areal productivity, provides a quantitative and standardized method for comparing methodologies. The results clearly indicate that sophisticated control strategies are essential to manage the complex interactions and strong diurnal forcing typical of ORPs, achieving performance metrics significantly superior to classical approaches.

This benchmark can serve as a crucial foundation for future control research in bioprocess systems and environmental engineering. Key lines of future work include:

\begin{itemize}
    \item Advanced multivariable and coupled control: developing and testing robust multivariable control strategies that explicitly account for the strong coupling between thermal, physicochemical, and biological states (e.g., how $\text{CO}_2$ injection affects pH and growth, or how temperature affects gas solubility and biomass kinetics).
    \item Alternative optimization formulations: investigating variations of the EMPC strategy, such as modifying the cost function to reward biomass growth instead of solely harvested biomass, or employing longer prediction horizons to fully capture medium-term economic benefits and sustainable operation trade-offs.
    \item Handling uncertainty and disturbance rejection: utilizing the benchmark's realistic scenario data (irradiance, wind, temperature) to develop control algorithms that explicitly manage the significant and non-linear effects of multi-rate meteorological disturbances.
    \item Fault detection and system identification: exploiting the high-fidelity dynamic model to test advanced system identification and fault diagnosis methods.
\end{itemize}

The benchmark is thus positioned to facilitate methodological progress and bridge the gap between control theory and real-world ORP operation.

\begin{table*}[ht]
    \centering
    \begin{tabular}{lcccc}
        \toprule
        & Player 1 & Player 2 & Player 3 & Player 4 \\
        \midrule
        \multicolumn{5}{c}{\textbf{Control modes}} \\
        \midrule
        pH control mode              & On/Off & PI & PI & PI \\
        DO control mode              & On/Off & PI & PI & PI \\
        Temperature control mode     & No Control & On/Off & PI & PI \\
        Harvesting strategy          & Fixed 20\% & Fixed 20\% & Turbidostat & EMPC \\
        \midrule
        \multicolumn{5}{c}{\textbf{Dynamic cost functions ($J$})} \\
        \midrule
        $J_{\text{pH}}$              & 1.0000  & 0.2143 & 0.2160 & 0.2226 \\
        $J_{\text{DO}}$              & 1.0000  & 0.7800 & 0.7535 & 0.7713 \\
        $J_{\text{Temp}}$     & 1.0000  & 0.4001 & 0.2021 & 0.2201 \\
        \textbf{$J_{\text{avg}}$ }& \textbf{1.0000}  & \textbf{0.4648} & \textbf{0.3905} & \textbf{0.4047} \\
        \midrule
        \multicolumn{5}{c}{\textbf{Key Performance Indicators (KPIs)}} \\
        \midrule
        Total air injected [L]                      & 2171500.00 & 2407410.41 & 2418712.22 & 2659724.47 \\
        Total CO$_2$ injected [L]                   & 112800.00  & 125056.98  & 127147.18  & 132840.88 \\
        Total biomass produced [g]                  & 23366.29   & 24828.80   & 25407.06   & 27376.60 \\
        Productivity per unit area [g m$^{-2}$ day$^{-1}$]    & 16.23      & 17.24      & 17.64      & 19.01 \\
        Harvested amount [g]                        & 22765.03   & 23716.49   & 25461.93   & 27071.85 \\
        Harvested per unit area [g m$^{-2}$ day$^{-1}$]       & 15.81      & 16.47      & 17.68      & 18.80 \\
        Production yield ratio [\%]                 & 97.43      & 95.52      & 100.22     & 98.89 \\
        Relative biomass accumulation [\% of initial]
                                                    & 9.57       & 17.71      & -0.87      & 4.85 \\
        \bottomrule
    \end{tabular}
    \caption{Quantitative results for the four simulated players. Cost functions $J$ are normalized with respect to Player 1}
    \label{tab:indices}
\end{table*}

\section*{Acknowledgments}
This work has been financed by the following projects: PID2023-150739OB-I00 project financed by the Spanish Ministry of Science and also by the European Union (Grant agreement IDs: 101060991, REALM;101146861, NIAGARA; 101214199, ALLIANCE)”). 

\bibliographystyle{elsarticle-harv} 
\bibliography{cas-refs}

%%%%%%%%%%%%%%%%%%%%%%%%%%%%%%%%%%%%%%%%%%%%%%%%%%%%
\section*{Author Note}

During the preparation of this work, the authors used \textit{ChatGPT (OpenAI)} as an assistance tool for text generation and technical editing. After using this tool, the authors thoroughly reviewed, modified, and validated the content, and they take full responsibility for the final version of the manuscript.

%%%%%%%%%%%%%%%%%%%%%%%%%%%%%%%%%%%%%%%%%%%%%%%%%%%%

\appendix
\section{Description of the \texttt{results} structure}
\label{app:results-structure}

\begin{itemize}
  \item \textbf{Time and references}
  \begin{itemize}
    \item \texttt{results.t}: time vector of the simulation horizon [s].
    \item \texttt{results.refs}: structure or matrix collecting the reference
          trajectories used by the controllers (e.g., setpoints for pH, DO, and
          temperature).
  \end{itemize}

  \item \textbf{Measured and reported process variables}
  \begin{itemize}
    \item \texttt{results.pH}: pH signal (dimensionless).
    \item \texttt{results.DO}: dissolved oxygen in percent saturation [\%].
    \item \texttt{results.T}: bulk reactor temperature [\si{\celsius}].
    \item \texttt{results.X\_gL}: biomass concentration expressed in
          \si{g.L^{-1}}.
    \item \texttt{results.Depth}: culture depth [m], computed from volume and
          geometry.
  \end{itemize}

  \item \textbf{Control commands and effective flow rates}
  \begin{itemize}
    \item \texttt{results.Qair\_cmd}: air injection command (controller output),
          before actuator limits are applied.
    \item \texttt{results.QCO2\_cmd}: CO$_2$ injection command.
    \item \texttt{results.Qair\_del}: delivered air flow rate after saturation
          and actuator constraints [m$^3$\,s$^{-1}$].
    \item \texttt{results.QCO2\_del}: delivered CO$_2$ flow rate
          [m$^3$\,s$^{-1}$].
    \item \texttt{results.Qd}: dilution inflow effectively used in the model
          [m$^3$\,s$^{-1}$].
    \item \texttt{results.Qh}: harvesting outflow effectively used in the model
          [m$^3$\,s$^{-1}$].
  \end{itemize}

  \item \textbf{Cumulative gas usage and biomass metrics}
  \begin{itemize}
    \item \texttt{results.cum\_air\_L}: cumulative volume of air injected over
          time [L].
    \item \texttt{results.cum\_CO2\_L}: cumulative volume of CO$_2$ injected
          [L].
    \item \texttt{results.cum\_harv\_g}: cumulative harvested biomass [g].
    \item \texttt{results.total\_air\_L}: total air consumption over the entire
          horizon [L].
    \item \texttt{results.total\_CO2\_L}: total CO$_2$ consumption [L].
  \end{itemize}

  \item \textbf{Key performance indicators (KPIs)}
  \begin{itemize}
    \item \texttt{results.gain\_g}: net biomass gain between the initial and
          final time [g].
    \item \texttt{results.acumm\_rel}: relative biomass accumulation index
          (dimensionless), comparing final and initial biomass inventories.
    \item \texttt{results.prod\_g}: total biomass produced (gross production)
          [g].
    \item \texttt{results.prod\_areal\_gm2\_day}: areal productivity
          [g\,m$^{-2}$\,day$^{-1}$].
    \item \texttt{results.harv\_total\_g}: total harvested biomass [g].
    \item \texttt{results.harv\_frac}: fraction of produced biomass that is
          effectively harvested (dimensionless).
    \item \texttt{results.harv\_prod\_areal\_gm2\_day}: areal productivity
          associated with the harvested biomass
          [g\,m$^{-2}$\,day$^{-1}$].
  \end{itemize}

  \item \textbf{Biological rates and limitation factors}
  \begin{itemize}
    \item \texttt{results.mu\_I}: light limitation factor $\mu_I$.
    \item \texttt{results.mu\_T}: temperature limitation factor $\mu_T$.
    \item \texttt{results.mu\_pH}: pH limitation factor $\mu_{\mathrm{pH}}$.
    \item \texttt{results.mu\_DO}: DO inhibition term $\mu_{DO}$.
    \item \texttt{results.P}: gross photosynthetic rate $P$
          [\si{d^{-1}} or s$^{-1}$, depending on model parametrization].
    \item \texttt{results.mu}: net specific growth rate $\mu_g$
          [\si{d^{-1}} or s$^{-1}$].
    \item \texttt{results.m}: maintenance/respiration rate $m$
          [\si{d^{-1}} or s$^{-1}$].
  \end{itemize}

  \item \textbf{Carbonate system and cations}
  \begin{itemize}
    \item \texttt{results.DIC}: dissolved inorganic carbon concentration
          [mol\,m$^{-3}$].
    \item \texttt{results.Cat}: strong cation concentration [mol\,m$^{-3}$].
    \item \texttt{results.HCO3}: bicarbonate concentration HCO$_3^-$
          [mol\,m$^{-3}$].
    \item \texttt{results.CO3}: carbonate concentration CO$_3^{2-}$
          [mol\,m$^{-3}$].
    \item \texttt{results.CO2}: dissolved CO$_2$ concentration
          [mol\,m$^{-3}$].
  \end{itemize}

  \item \textbf{Heat-exchanger substructure}
  \begin{itemize}
    \item \texttt{results.HX.Qw\_m3s}: water flow rate through the spiral heat
          exchanger [m$^3$\,s$^{-1}$].
    \item \texttt{results.HX.Tin\_C}: inlet water temperature to the heat
          exchanger [\si{\celsius}].
    \item \texttt{results.HX.Tout\_C}: outlet water temperature from the heat
          exchanger [\si{\celsius}].
    \item \texttt{results.HX.UA\_WK}: overall heat-transfer coefficient times
          area, $UA$ [W\,K$^{-1}$].
    \item \texttt{results.HX.Q\_W}: instantaneous heat flux exchanged with the
          culture, $Q_{\mathrm{HX}}$ [W].
    \item \texttt{results.HX.limits}: structure containing actuator limits for
          the heat-exchanger loop:
          \texttt{Qw\_min}, \texttt{Qw\_max}, \texttt{Tin\_min}, and
          \texttt{Tin\_max}.
  \end{itemize}
\end{itemize}
\end{document}